\documentclass[%
 reprint,
superscriptaddress,
 amsmath,amssymb,
 aps,prl
]{revtex4-2}
\usepackage{graphicx}
\usepackage{dcolumn}
\usepackage{bm}

\usepackage{xcolor}

\begin{document}

\preprint{APS/123-QED}

\title{Revealing Actual Viscoelastic Relaxation Times in Capillary Breakup }

\author{Nan Hu}
\email{nh0529@princeton.edu}
\author{Jonghyun Hwang}
\affiliation{%
Department of Mechanical and Aerospace Engineering, Princeton University, NJ 08544, USA
}%

\author{Tachin Ruangkriengsin}
\affiliation{%
Program in Applied and Computational Mathematics, Princeton University, NJ 08544, USA
}%
\author{Howard A. Stone}%
 \email{hastone@princeton.edu}
\affiliation{%
Department of Mechanical and Aerospace Engineering, Princeton University, NJ 08544, USA
}%

\date{\today}

\begin{abstract}
We use experiments and theory to elucidate the size effect in capillary breakup rheometry, where pre-stretching in the visco-capillary stage causes the apparent relaxation time to be consistently smaller than the actual value. We propose a method accounting for both the experimental size and the finite extensibility of polymers  to extract the actual relaxation time. A phase diagram characterizes the expected measurement variability and delineates scaling law conditions. The results  refine capillary breakup rheometry for viscoelastic fluids and advance the understanding of breakup dynamics across scales.

\end{abstract}

\maketitle

Capillary thinning and breakup of viscoelastic liquid threads are ubiquitous in natural environments, biological tissues, and industrial applications \cite{Eggers_1997,mckinley2005visco,Keshavarz_2016}. Unlike viscous liquids, the capillary breakup of viscoelastic liquids exhibits distinct behaviors such as the exponential decay of filament thinning \cite{Entov_1997,Amarouchene2001,Eggers_2020,Deblais_2020} and the formation of satellite droplets on a string \cite{Wagner2005,bhat2010formation,Deblais_2018,Turkoz_2018}, which are attributed to the elastic forces from polymeric stresses. Consequently, capillary breakup rheometry, which leverages these dynamic characteristics, has been a method widely used to measure viscoelastic properties, and is especially promising  for weakly-elastic fluids \cite{rodd2005capillary} and biopolymers such as DNA \cite{Ingremeau2013,Calabrese2024}, whose properties are difficult to measure using common rheometric methods. Current well-developed experimental techniques include stretching a liquid bridge between two plates (known as CaBER) \cite{Anna2001,McKinley2002,Aisling2024,Gaillard_2024}, jetting \cite{Eggers2008,ARDEKANI_2010,sharma2015rheology,Keshavarz_2015}, dripping \cite{Rajesh2022,Bazazi2023}, and dripping-onto-substrate (DoS) \cite{dinic2017pinch,Dinic_2017,jimenez2018extensional,Zinelis_2024}.

Typically, the capillary breakup of viscoelastic fluids is governed by the interplay of inertial, viscous, elastic, and surface tension forces. Based on the sequence of events during the breakup of droplets or liquid bridges, the dynamics can be categorized into the visco-capillary (VC) or inertial-capillary (IC) stage, the elasto-capillary (EC) stage, and the final visco-elasto-capillary (VEC) stage \cite{dinic2019macromolecular}. By assuming infinite extensibility  of the polymer chains ($b=\infty$) with a  (longest) relaxation time $\lambda$, the evolution of the neck diameter $D(t)$ for the EC stage is well-known as 
\begin{equation}
D(t)/D_1 = \left(GD_1/2\gamma\right)^{1 / 3} \exp \left[-\left(t-t_1\right) / 3 \lambda\right], 
\label{exp_decay}
\end{equation}
where $G$ is the modulus of the viscoelastic fluid, $D_1$ is the initial diameter of the liquid filament, $\gamma$ is the surface tension, and $t_1$ is the time of onset of the EC stage \cite{Entov_1997}.
This formula has been adopted in numerous studies to interpret the experimental results of the variation of the filament diameter  $D(t)$ during the EC stage. Thus, one obtains an estimate of the apparent relaxation time $\lambda_e$ for various kinds of polymer solutions by fitting the exponential decay \cite{Anna2001,rodd2005capillary,arnolds2010capillary,dinic2015extensional,dinic2017pinch,Dinic_2017,jimenez2018extensional,dinic2019macromolecular,Zinelis2024,Arxiv_prx}.

\begin{figure*}
    \centering
    \includegraphics[width=\textwidth]{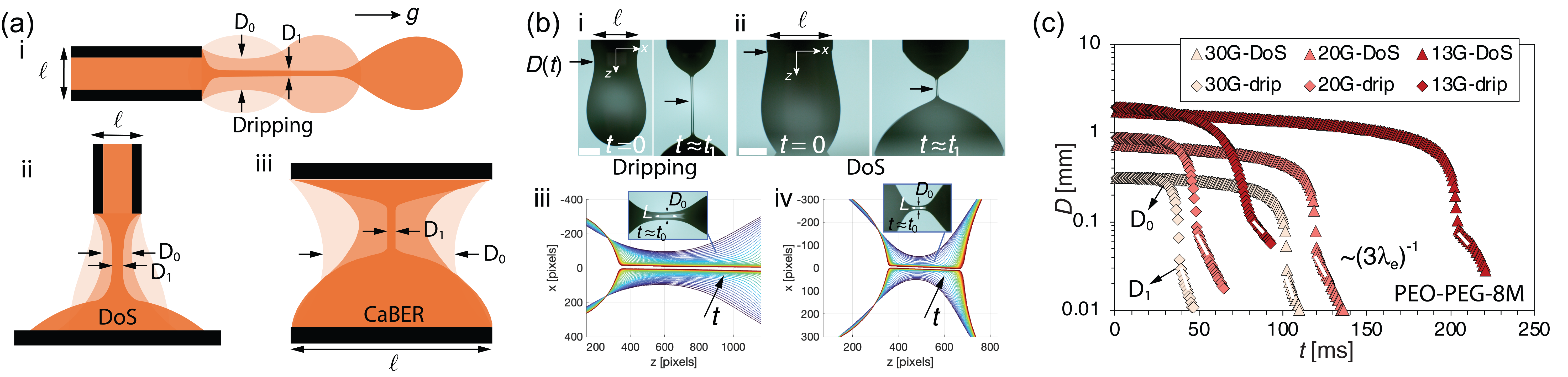}
    \caption{Capillary breakup rheometry. (a) Schematic illustrations of liquid filament breakup for different experimental configurations: (i) dripping, (ii) dripping-onto-substrate (DoS), and (iii) a liquid bridge between two plates (classic CaBER). The liquid is initially held quasi-steady by a nozzle or plates with diameter $\ell$. The minimum neck diameter of the liquid, $D(t)$, evolves over time from the $D_0$ to $D_1$, where subscripts 0 and 1 mark the onset of the VC and EC stages, respectively. (b) Representative measurements for a PEO-PEG-8M solution  using (i) dripping and (ii) DoS methods with a 13G nozzle ($\ell = 2.4 \, \mathrm{mm}$). (iii) and (iv) are the extracted evolutions of interface shapes with increasing time $t$ corresponding, respectively, to dripping and DoS, where $D_0$ is estimated by taking the filament aspect ratio $1\lesssim L/D_0 \lesssim 3$.  
    Scale bars represent 1 mm. (c) Representative datasets showing the variation of $D(t)$ for different $\ell \in \{0.3, 0.9, 2.4\} \, \mathrm{mm}$ 
    and methods (different symbols). Distinct slopes of $\log_{10}D$ in the EC stage for varying $\ell$ indicate a size-dependent apparent relaxation time $\lambda_e$ predicted by the slope $\sim (3\lambda_e)^{-1}$.} 
    \label{Fig1}
\end{figure*}
However, existing studies have found that pre-stretching of the polymers or filament thinning during the IC/VC stage significantly influences the exponential decay in the subsequent EC stage~\cite{Calabrese2024,Aisling2024,Gaillard_2024,Rajesh2022,Zinelis2024}. The sensitivity of $\lambda_e$ to these effects can be attributed to several experimental parameters, including the initial separation rate of the plates~\cite{Aisling2024}, the initial diameter $D_0$ of the liquid bridge~\cite{Gaillard_2024} in the CaBER method, nozzle diameter $\ell$ in the dripping method (where $\lambda_e \propto \ell^{8/9}$ has been observed)~\cite{Rajesh2022}, and the polymer extensibility parameter $b$ in the DoS method~\cite{Zinelis2024}. 
This parameter dependence introduces strong sensitivity of the extracted relaxation time $\lambda_e$ to initial conditions and inevitable pre-stretching effects in capillary breakup rheometry.
{
While several studies have reported such dependence~\cite{Rajesh2022,Gaillard_2024,Zinelis2024}, no work has yet directly identified the actual relaxation time $\lambda$ from breakup measurements, nor established a general scaling law relating $\lambda_e$ to $\ell$ or $D_0$. 
In contrast, a recent study~\cite{Arxiv_prx} reported no systematic dependence of $\lambda_e$ on CaBER plate size across four polymer solutions with varying concentration and molecular weight. This highlights the need for a unified framework to account for the size-dependent variability of $\lambda_e$ observed in different experimental systems.}

In this Letter we demonstrate that size-dependent apparent relaxation times, $\lambda_e$, obtained from capillary breakup processes of varying scales, can be used to estimate the actual relaxation time, $\lambda$, while simultaneously determining the extensibility, $b$, representative of the polymer solution. In particular, accounting for finite extensibility $b$, we show that pre-stretching during the VC stage  
shortens the exponential decay in the elasto-capillary stage, leading to $\lambda_e<\lambda$, and that this effect can be dependent on the size of the experimental setups. Through experiments and numerical solutions, we unravel this quantitative relationship, enabling the determination of the actual relaxation time $\lambda$ from both our measurements and those reported in recent studies. 

The capillary breakup processes considered here include dripping, DoS, and 
CaBER (Fig.~\ref{Fig1}a (i, ii, iii)). In all cases, the liquid initially is slowly stretched in a quasi-static manner within devices of characteristic size $\ell$. Under the action of capillary forces driving the Rayleigh-Plateau instability, the liquid rapidly contracts from the initial diameter $D_0$ of the VC stage to the initial diameter $D_1$ of the EC stage. In our experiments, we used needles with varying outer diameter $\ell \in\{0.3,0.5,0.9,1.8,2.4\}$ mm (corresponding to 30G, 25G, 20G, 15G, and 13G needles) to conduct tests using the dripping and DoS methods. Given that the influence of the size of the CaBER method has  been tested in Ref.~\cite{Gaillard_2024,Arxiv_prx}, we also used their data of the PEOvis and PS solution (see \cite{SupportingInformation}, \S I and Fig.~S1) for our analysis. 

Three kinds of polymer solutions were employed in our experiments: PIB-PB (polyisobutylene dissolved in a polybutene-mineral oil mixed solvent), PS-DOP (polystyrene dissolved in the organic solvent dioctyl phthalate), and PEO-PEG (polyethylene oxide dissolved in a polyethylene glycol-water solution), whereas PIB-PB-0.3 and PIB-PB-0.02 denote different PIB concentrations; PEO-PEG-8M and PEO-PEG-1M identify different molecular weights of PEO. These latter fluids can be regarded as Boger fluids with constant viscosity (Fig. S2 \cite{SupportingInformation}), ensuring that additional effects from viscosity variations with extension rate are eliminated. Detailed preparation procedures and physical property measurements, such as viscosity $\eta_0$, surface tension $\gamma$, storage modulus $G'$, and loss modulus $G''$, are provided in \cite{SupportingInformation}~\S II and summarized in Table S1. 
{The polymer concentrations $c$ used in this study fall within the dilute regime, with $c^* > c \gtrsim 0.1c^*$, and are all significantly higher than the minimal concentrations required for EC stage \cite{Clasen2006,CampoDeao2010}. These include $c_{\mathrm{min}} \sim c^*/b$ and $c_{\mathrm{low}} \sim c^*/b^{0.75}$, which represent the thresholds for the emergence of the EC regime \cite{Clasen2006} and for visibly observable exponential thinning \cite{CampoDeao2010}, respectively.}

Capillary breakup tests were conducted inside a transparent chamber to shield the system from air disturbances, with a liquid reservoir included to minimize the effects of evaporation (see \cite{SupportingInformation}~\S III). High-speed imaging was conducted using a camera (VEO 640L, Phantom) equipped with microscope objectives (Nikon) of 2X, 5X, or 10X magnifications to accommodate varying nozzle diameters. A syringe pump (PHD 2000, Harvard Apparatus) controlled the flow rate at $Q = 1-10 \, \mu\mathrm{L/min}$, ensuring a quasi-static droplet for dripping. Taking the PEO-PEG-8M solution as an example, representative snapshots from the dripping and DoS tests, captured at a resolution of $2560 \times 1600$ pixels and 1500 frames per second (fps), are shown in Fig.~\ref{Fig1}(b-i, ii), respectively. 
The liquid surface was analyzed using a custom MATLAB image-processing algorithm to track the gas-liquid interface as depicted in Fig.~\ref{Fig1}(b-iii, iv), in order to obtain the evolution of  the minimum neck diameter $D(t)$ (Fig.~\ref{Fig1}c). Unlike the CaBER method, where the instability can be triggered in a controlled manner by adjusting the plate spacing to directly determine \( D_0 \), we propose that in the dripping and DoS methods, \( D_0 \) should be determined by ensuring \( 1\lesssim L/D_0 \lesssim 3 \) (Fig.~\ref{Fig1}b-iii, iv, detailed discussion seen in \cite{SupportingInformation} \S V). Each case was repeated at least three times to confirm data reproducibility. The  results for the other solutions are provided in Fig.~S4 ~\cite{SupportingInformation}.
\begin{figure}
    \centering \includegraphics[width=0.8\columnwidth]{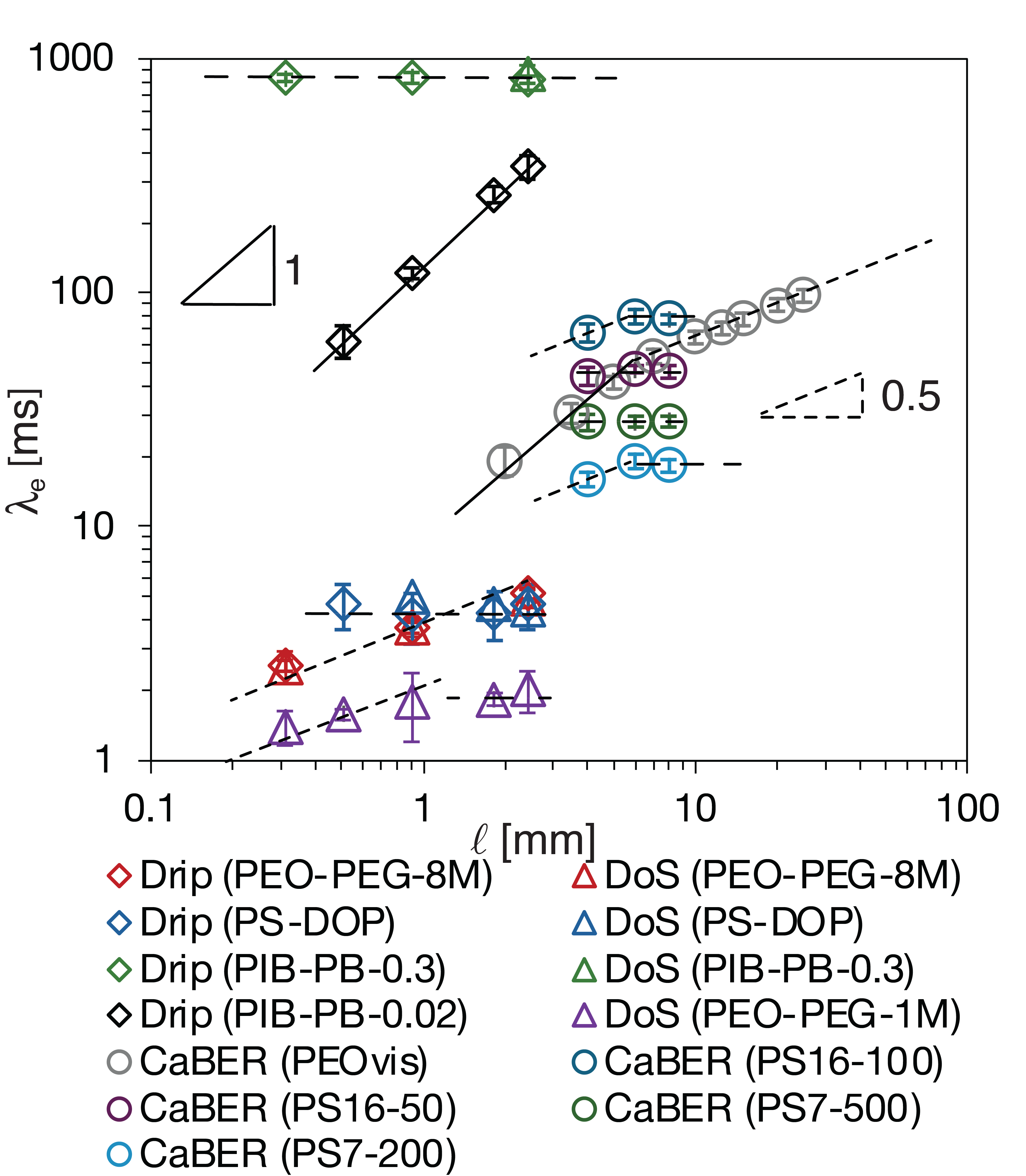}
    \caption{Dependence of apparent relaxation time $\lambda_e$ on apparatus size $\ell$ for different Boger fluids using specific methods: PEO-PEG, PS-DOP, PIB-PB in this work by dripping and/or DoS, PEOvis (Ref.~\cite{Gaillard_2024}) {and PS16/PS7 (Ref.~\cite{Arxiv_prx})}
    by CaBER.}
    \label{Fig2}
\end{figure}

The transition in the $D(t)$ curves and the appearance of a uniform filament (Fig.~\ref{Fig1}b-i, ii) allow the determination of the 
initial diameter $D_1$ for the EC stage. Using Eq. (\ref{exp_decay}), a fit following this point (white lines in Fig.~\ref{Fig1}c) yields the value of $\lambda_e$. The results show that $\lambda_e$ increases with the nozzle diameter $\ell$, though no significant difference is observed between the $\lambda_e$ values obtained from the dripping and DoS methods. We conducted a comprehensive comparison of the apparent relaxation times $\lambda_e$ for different viscoelastic fluids, nozzle diameters $\ell$, and testing methods (Fig.~\ref{Fig2}). We observed that PEO-PEG, PIB-PB-0.02, and PEOvis solutions exhibit a size dependence of $\lambda_e$, showing a  
variation from $\sim \ell$ to $\sim\ell^{0.5}$ with increasing $\ell$ for one fluid. In contrast,  $\lambda_e$ values of the PS-DOP and PIB-PB-0.3 solution remain nearly constant.

\begin{figure}
    \centering \includegraphics[width=\columnwidth]{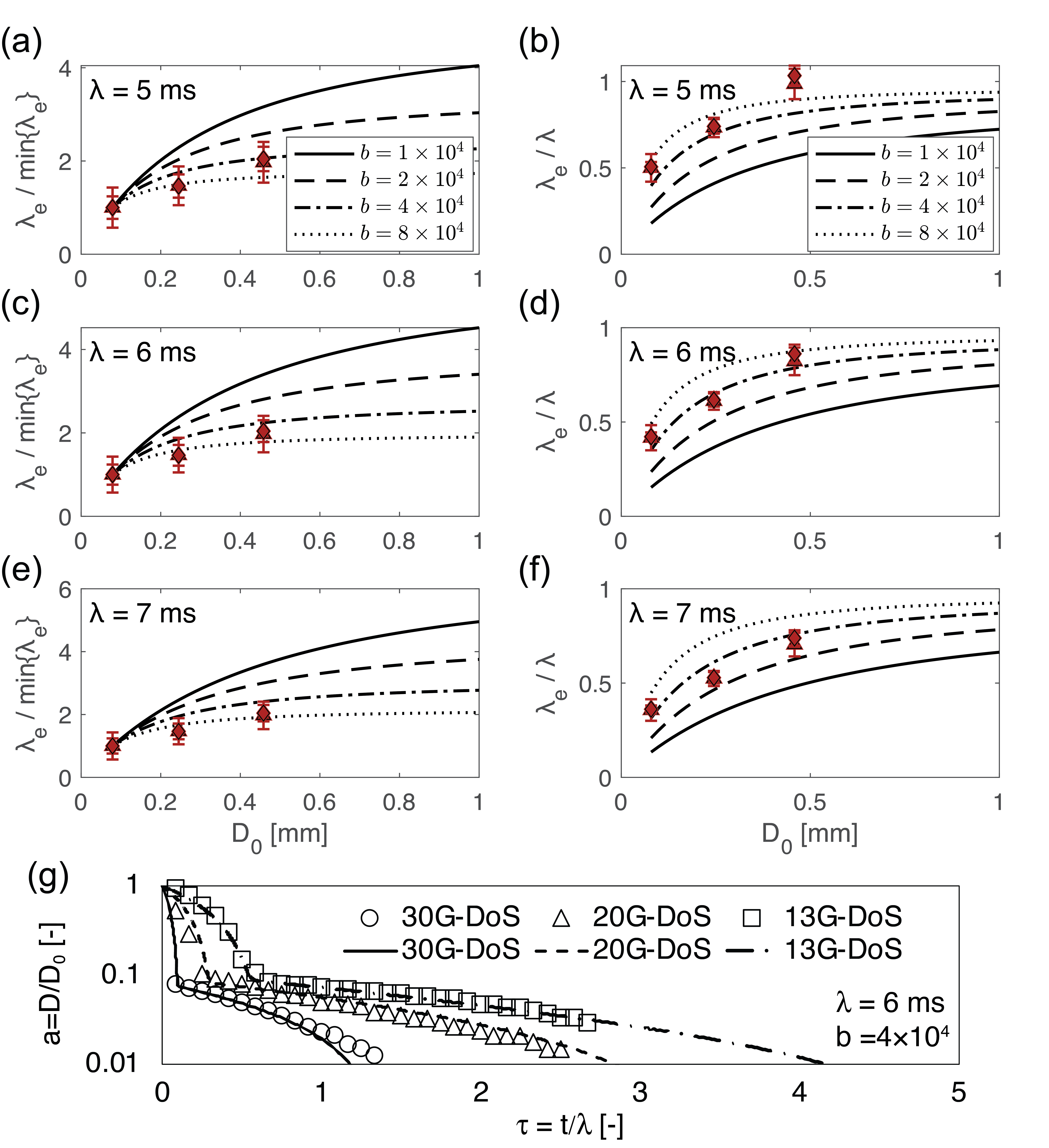}
    \caption{Prediction of the relaxation time ratio $\lambda_e/\mathrm{min}\{\lambda_e\}$ (a, c, e) and relative ratio $\lambda_e / \lambda$ (b, d, f) versus the initial liquid diameter $D_0$ for the PEO-PEG-8M solution, assuming $\lambda =$  5 ms (a, b),  6 ms (c, d), and 9 ms (e, f), with varying extensibility $b \in \{2\times10^4, 4 \times 10^4, 8 \times 10^4\}$. The symbols represent experimental results of $\lambda_e$ obtained from the dripping and DoS methods, suggesting a good approximation of $\lambda \approx 6$ ms and simultaneously indicating $b \approx 4\times10^4$ for the PEO-PEG solution. (g) Comparison of experimental and numerical results using $(\lambda,b)=(6~\mathrm{ms},4\times10^4)$ for three different diameter needles.}     \label{Fig3}
\end{figure}
We believe that viscoelastic fluids formed from nearly monodisperse polymer solutions should exhibit a constant intrinsic (longest) relaxation time $\lambda$, independent of testing conditions; indeed, this is the most common description of these solutions. 
To uncover the underlying dependence of $\lambda_e$ on experimental size $\ell$, we employed a zero-dimensional model based on the FENE-P model to predict the evolution of the dimensionless filament diameter $a(\tau)=D(\tau)/D_0$, according to equations for mass and momentum balances (subscripts $s$ and $p$ denote, respectively, solvent and polymer), and microstructure evolution along the axial ($z$) and radial ($r$) directions, 
\begin{subequations}
    \begin{equation}
    \frac{Ec}{a} = \frac{3\beta}{1-\beta}E + f(A_{zz}-A_{rr})
    \label{full_model_a}
    \end{equation}
    \begin{equation}
    \partial_\tau A_{zz}   = \left(2E-f\right)A_{zz}+1
    \end{equation}
    \begin{equation}
  \partial_\tau A_{rr}   = \left(-E-f\right)A_{rr}+1
    \end{equation}
    \begin{equation}
    \partial_\tau a = -\frac{aE}{2},
    \end{equation}
\label{FilamentThinninEqnSet}
\end{subequations}
where we have introduced dimensionless time $\tau = t/\lambda$, elasto-capillary number $Ec = 2\lambda\gamma/\eta_pD_0$, solvent viscosity ratio $\beta = \eta_s/\eta_0$, dimensionless (time-varying) extensional strain rate $E = -2\lambda D^{-1}\frac{dD}{dt}$,
the conformation tensor $\boldsymbol{A}$ with $rr$ and $zz$ components, and the finite extensibility function $f=(b-3)/(b-A_{zz}-2A_{rr})$.
By comparing simplified analytical solutions (see \cite{SupportingInformation} \S IV-B1, B2), approximate solutions (\cite{SupportingInformation} \S IV-B3), and an analysis of the full model   (\cite{SupportingInformation} \S IV-C1) under different assumptions, we demonstrated that the evolution is sensitive to the initial conformation $\boldsymbol{A}(t_1)$ for the EC stage. Only by solving the full model numerically can the variations in the VC, EC, and VC-EC transitions be  captured accurately (see \cite{SupportingInformation} \S IV).

With the initial conditions $a=A_{rr}=A_{zz}=1$, 
Eqs. (\ref{FilamentThinninEqnSet}) can be solved numerically (\cite{SupportingInformation}~\S IV-C2) using MATLAB's ODE15s solver to yield the temporal evolution of the  dimensionless diameter $a(\tau)$, conformation $A_{zz}$ and $A_{rr}$, and extensional strain rate $E(\tau)$ (typical results are shown in Fig.~\ref{Fig3} for $a(\tau)$; see also Fig.~S10 for all variables).
Based on the numerical results, we calculate the numerically predicted $\lambda_e / \lambda = \max\left\{ {2}/{(3E(t))}\right\}$ for $t\geq t_1$, which corresponds to the early period of the EC stage (see \cite{SupportingInformation}, Fig.~S10d), under various parameter choices of $\beta$, $Ec$, and $b$. This further allows the construction of a phase diagram of $\lambda_e / \lambda$ as a function of the three parameters (see \cite{SupportingInformation}, Fig.~S10e). Notably, every fixed $\beta$ enables the phase diagram to be approximated with  a two-dimensional representation in terms of $Ec$ and $b$ (see \cite{SupportingInformation}, Fig.~S10f). 

For a specific viscoelastic liquid, as properties $\eta_p$, $\beta$ and $\gamma$ are measurable and size $D_0$ is controlled, we can assume a true relaxation time $\lambda$ for determining $Ec$ in Eq. (\ref{full_model_a}), and explore a wide range of $b$ values in the numerical calculations. 
Taking the PEO-PEG solution as an example, numerical predictions of the relative ratio $\lambda_e/\mathrm{min}\{\lambda_e\}$  (Fig.~\ref{Fig3}a, c, e) and $\lambda_e / \lambda$ (Fig.~\ref{Fig3}b, d, f) for assumed $\lambda \in \{5,6,7\} \, \mathrm{ms}$ and $b \in \{10^4,2\times10^4,4\times10^4,8\times10^4\}$ are compared with experimental results (symbols in Fig.~\ref{Fig3}). 
The trend of  $\lambda_e/\mathrm{min}\{\lambda_e\}$ for 
 different $D_0$ values provides a significant indicator for comparison with model predictions because both $\lambda_e/\lambda$ and $\lambda_e/\mathrm{min}\{\lambda_e\}$ are uniquely determined for fixed $Ec$ and $\beta$.

The comparisons of the two trends indicate that when $\lambda\approx6 \, \mathrm{ms}$, the numerical predictions align well with the experimental results (dash-dot curves in Fig. \ref{Fig3}g). Furthermore, at the same time this analysis  determines  $b \approx 4\times10^4$. We emphasize that with this approach, only $\lambda$ needs to be adjusted for comparison, while $b$ is  determined through the matching of the two trends. 
The comparison of experimental and numerical $a(\tau)$ shown in 
Fig.~\ref{Fig3}g suggests a good estimation for $\lambda$ and $b$ has been achieved. The same procedure was used to find a good agreement with the experimental data from \cite{Gaillard_2024} (Fig. S11 in \cite{SupportingInformation}).

\begin{figure}
    \centering \includegraphics[width=\columnwidth]{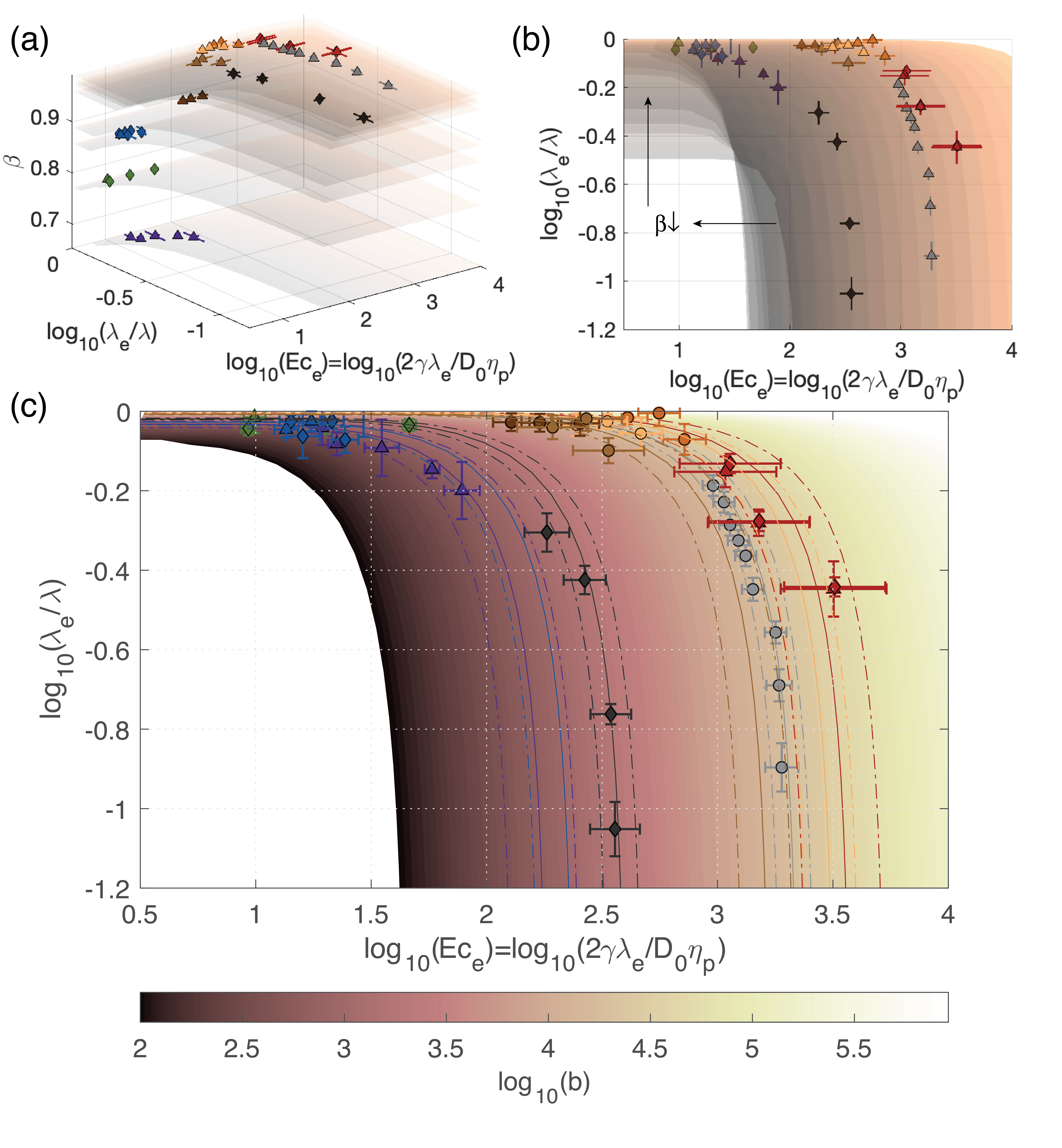}
    \caption{Dimensionless map of the relaxation time ratio $\lambda_e / \lambda$ versus the measurable elasto-capillary number $Ec_e =  2\gamma \lambda_e/(D_0 \eta_p)$ for various unknown extensibility values $b \in [10^2, 10^6]$. (a) 3D maps for different $\beta$ of a specific solution. (b) Top view of 3D maps. (c) Collapsed 2D map for easy comparison. The symbols represent experimental results (red: PEO-PEG-8M, purple: PEO-PEG-1M, green: PIB-PB-0.3, black: PIB-PB-0.02, blue: PS-DOP) from current work, Ref.~\cite{Gaillard_2024} (gray: PEOvis), {and Ref.~\cite{Arxiv_prx} (light/dark orange: PS16-50/100, light/dark brown: PS7-200/500)}. Solid lines are obtained from fitting, with deviation ranges enclosed by dash-dot lines, suggesting the estimations.
    } 
    \label{Fig4}
\end{figure}
Next, we analyzed the experimental data presented in Fig.~\ref{Fig2} and constructed a dimensionless phase diagram (note the logarithmic axes) of $\lambda_e / \lambda$ as a function of $Ec_e$, $b$ and $\beta$ (Fig.~\ref{Fig4}a), where $Ec_e =  2 \gamma \lambda_e/(D_0 \eta_p)$, with each variable being directly measurable from experiments. Although different $\beta$ values affect the map, particularly for small {$b$}, indicated in Fig.~\ref{Fig4}b, we can construct the phase diagram with all experimental data as shown in Fig.~\ref{Fig4}c; uncertainties in all measurable parameters were considered to determine the true relaxation time $\lambda$ and the range of extensibility $b$ for each viscoelastic fluid tested. 

For the PEO-PEG-8M solution, the relaxation time is estimated as $\lambda \approx 6 \, \mathrm{ms}$, with $b = 4 \times 10^{4\pm0.2}$. For the PEOvis solution, $\lambda \approx 150 \, \mathrm{ms}$ and $b = 2\times10^{4\pm0.1}$, consistent with the results from Ref.~\cite{Gaillard_2024}, where $b \approx 2\times 10^4$ was estimated using an additional model. This agreement reinforces the validity of our approach. Furthermore, the larger $b$ for PEO-PEG-8M compared to PEOvis aligns with their respective molecular weights, $M_w \sim 8\times 10^6$ in our experiments versus $M_w \sim 4\times 10^6$ in Ref.~\cite{Gaillard_2024}, given similar solvent mixtures (see \cite{SupportingInformation}, Tab.~S1). This observation is further supported by the PEO-PEG-1M solution with $M_w\sim 1\times 10^6$, which yields $(\lambda, b) \approx (2.2~\mathrm{ms}, 7 \times 10^{2\pm0.2})$.

For the PIB-PB-0.3 solution, experimental uncertainties prevent a precise determination of whether $\lambda_e$ varies with $D_0$, suggesting that a reasonable estimation of $\lambda$ and $b$ is not readily attainable. However, utilizing the more dilute PIB-PB-0.02 solution, we estimate $(\lambda, b) \approx (700 \, \mathrm{ms}, 2 \times 10^{3\pm0.1})$. Given that both solutions contain the same polymer at different dilutions, it is reasonable to assume that $b$ remains unchanged. This leads to an inferred relaxation time of $\lambda \approx 850 \, \mathrm{ms}$ for PIB-PB-0.3. 

From the phase diagram (Fig. \ref{Fig4}), two distinct regions can be identified. When {{$\log_{10}(Ec_e^{-1} b) \gtrsim 1.5$, the experimentally measured $\lambda_e$ becomes independent of the experimental size, i.e. $\lambda_e\sim\ell^0$. However, this does not imply that the true relaxation time $\lambda$ is obtained. In this region, since $b$ cannot be experimentally fitted, an additional evaluation of $b$ is necessary to determine the specific value of $\lambda_e / \lambda$. In the region where $\log_{10}(\lambda_e / \lambda) \lesssim -0.6$, the slope of $\log_{10}(\lambda_e / \lambda)$ approaches infinity, indicating that $\lambda_e \sim D_0$. This is consistent with the scaling law $\lambda_e\sim \ell$ observed in Fig.~\ref{Fig2} when $\ell(D_0)$ becomes sufficiently small. The $\lambda_e\sim \ell^{0.5}$ scaling law observed in Fig.~\ref{Fig2} represents an empirical scaling behavior characterizing the transition between the two regions.

Thus, some  predictions can be made. For fluids with relatively low surface tension and strong viscoelasticity, such as PIB-PB-0.3 and PS-DOP, which exhibit $\log_{10}(Ec_e)\lesssim 1.5$ during capillary breakup, a size-independent exponential decay in the EC stage may occur. This implies that a nearly constant apparent relaxation time can be observed. In contrast, fluids with relatively high surface tension and low viscoelasticity, such as water-based dilute solutions like PEO-PEG and PEOvis, exhibit significant size dependence of the apparent relaxation time. 
The extensibility $b$ determines the  value of the critical $Ec_e$ at which this effect occurs, with smaller $b$ making this influence more readily triggered. Additionally, we speculate that in micro-scale capillary breakup processes (e.g., within biological tissues), where $Ec_e$ is extremely big, the linear dependence of $\lambda_e \sim D_0\sim\ell$ is  likely to occur.

In summary, our study reveals, through experiments and numerical solutions, how inevitable pre-stretching during the visco-capillary stage of capillary breakup influences the exponential decay in the elasto-capillary stage and the apparent relaxation time $\lambda_e$ obtained using classical formulations for viscoelastic fluids. Through a theoretical model, we demonstrate that the apparent relaxation time is always smaller than the actual relaxation time, i.e., $\lambda_e<\lambda$. 
We further identify that the size dependence of $\lambda_e$ is governed by the extensibility $b$, solvent viscosity ratio $\beta$, and the dimensionless number $Ec_e = 2 \gamma \lambda_e/(D_0 \eta_p)$. 
Large $b$ and small $Ec_e$ result in size-independent behavior, while under big $Ec_e$ conditions, the scaling $\lambda_e \sim D_0 \sim \ell$ emerges. 
We present a methodology for determining the actual relaxation time $\lambda$ by synthesizing a series of apparent relaxation times $\lambda_e$ measured under various size conditions. These findings provide a new perspective on accurately measuring viscoelastic properties using capillary breakup rheometry and offer insights into the size effects in the capillary breakup dynamics of viscoelastic fluids.
Looking forward, several factors need to be considered to address the measurement of more complex fluids, such as those with shear-thinning viscosity, fluids where the IC stage dominates instead of the VC stage.

\begin{acknowledgments}
We thank Evgeniy Boyko for valuable discussions. We thank the NSF-supported project (NSF-BSF: Explaining the Mismatch of Experiments and Simulations for Viscoelastic Flows, 2246791) and the Princeton Materials Research Science and Engineering Center (MRSEC, DMR-2011750) for partially funding this work. J.H. acknowledges the Kwanjeong Educational Foundation Graduate Fellowship for financial support.
\end{acknowledgments}

%

\end{document}


\preprint{APS/123-QED}

\title{Supporting Information for ``Revealing Actual Viscoelastic Relaxation Time in Capillary Breakup"}

\author{Nan Hu}
\email{nh0529@princeton.edu}
\author{Jonghyun Hwang}
\affiliation{%
Department of Mechanical and Aerospace Engineering, Princeton University, NJ 08544, USA
}%

\author{Tachin Ruangkriengsin}
\affiliation{%
Program in Applied and Computational Mathematics, Princeton University, NJ 08544, USA
}%
\author{Howard A. Stone}%
 \email{hastone@princeton.edu}
\affiliation{%
Department of Mechanical and Aerospace Engineering, Princeton University, NJ 08544, USA
}%


\renewcommand{\figurename}{Fig.}
\renewcommand{\thefigure}{S\arabic{figure}}
\renewcommand{\theequation}{S\arabic{equation}}

\maketitle

\tableofcontents
\section{I. Existed data of CaBER method}
\begin{figure*}
\includegraphics[width=\textwidth]{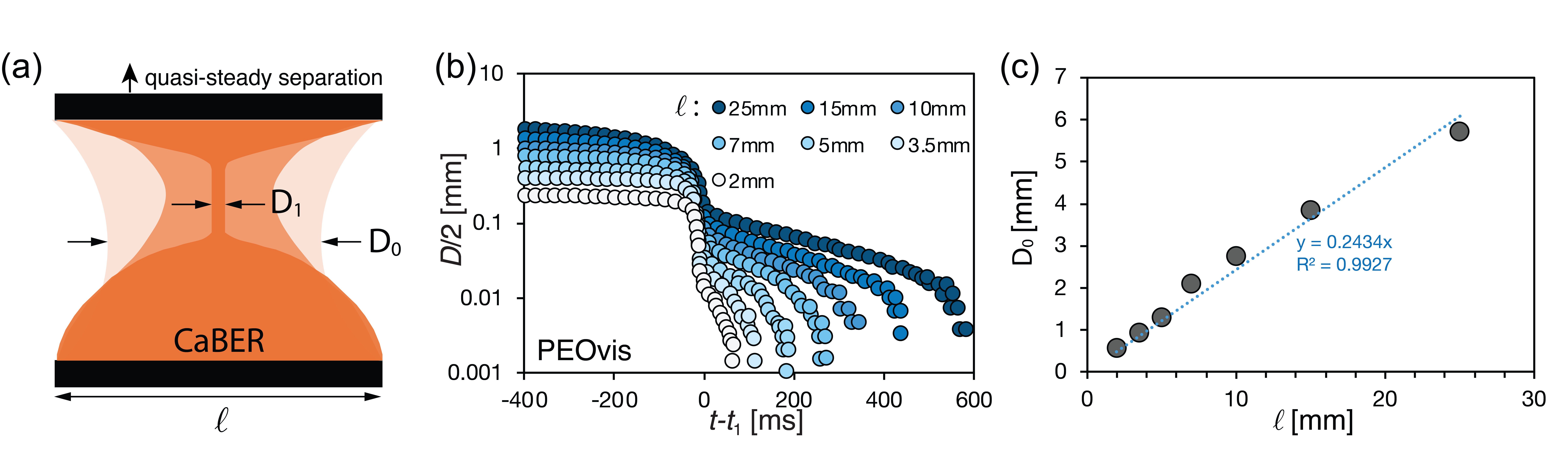}
\caption{\label{fig:FigureS1}  Experiments on the CaBER method with flat plates of varying diameter. (a) Schematic of the stepwise plate separation CaBER method, designed to maintain the liquid bridge in a quasi-steady state prior to capillary breakup. (b) Evolution of the liquid bridge neck radius $R=D/2$, as a function of time for different plate diameters $\ell$, using a PEOvis solution (reproduced from Ref.~\cite{Gaillard_2024} with the permission of American Physical Society). 
The time $t_1$ marks the onset of the elastic-capillary stage. 
(c) Dependence of the initial neck diameter $D_0$ on the plate diameter $\ell$ in experiments. The dotted line represents the linear fit $D_0 \approx 0.24\ell$ for PEOvis in the quasi-steady separation CaBER method.
}
\label{fig:S1}
\end{figure*}

Gaillard et al.~\cite{Gaillard_2024} measured the capillary breakup process of PEOvis solutions, i.e., polyethylene oxide polymer with $M_w=4\times10^6$ in a water-polyethylene glycol ($M_w=2\times10^4$) mixture, using the quasi-steady separating CaBER method between parallel plates of different diameters $\ell \in \{2, 3.5, 5, 7, 10, 15, 25\} \, \mathrm{mm}$ (see Fig.~\ref{fig:S1}a). The detailed measurement data were redrawn in Fig.~\ref{fig:S1}(b). Notably, the initial diameter of the liquid column, $D_0$, is not equal to $\ell$, but a linear relationship exists as shown in Fig.~\ref{fig:S1}(c) from their experimental results. The specific physical properties of the PEOvis solution are viscosity $\eta_0 = 0.248\pm0.001 \, \mathrm{Pa \, s}$, solvent viscosity $\eta_s = 0.246\pm0.001 \, \mathrm{Pa \, s}$ ($\beta=\eta_s/\eta_0=0.988\pm0.006$), and surface tension $\gamma = 56 \, \mathrm{mN/m}$.

{
Calabrese et al. \cite{Arxiv_prx} also have measured dilute polymeric solutions of polystyrene (PS) dissolved in dioctyl phthalate (DOP), with two average molecular weights $M_w\approx7\times10^6$ and $1.6\times10^7$ (labeled as PS7 and PS16). Four specific solutions named as PS16-100, PS16-50, PS7-500, and PS7-200 are adopted in the analysis, corresponding to concentrations of 100, 50, 500, and 200 ppm.  The related parameters are listed as: $\ell\in\{4,6,8\}$ mm, $\eta_s\approx57$ mPa s, $\gamma\approx31$ mN/m, and $\eta_p\approx\{2,2,7,3\}$ mPa s for PS16-100, PS16-50, PS7-500, and PS7-200. Extensibility $b=3.2\times10^4$ and $1.36\times10^4$ from \cite{Arxiv_prx} are adopted in the full model analysis.
}

\section{II. Viscoelastic Liquid Preparation, Characterization and Selection}
{PEO-PEG Solution Preparation:}  First, 60 wt.\% deionized water was weighed, followed by adding 40 wt.\% of polyethylene glycol (PEG, average $M_w = 8000$ g/mol, Aladdin). The mixture was stirred using a magnetic stirrer at room temperature for 24 hours until fully dissolved. Then, polyethylene oxide polymer (PEO, Sigma-Aldrich) was added and stirred for another 24 hours. Three different molecular weight PEOs were used including $M_w=8\times10^6, 1\times10^6$ and $1\times10^5$, denoted as 8 M, 1 M and 0.1 M, respectively. 
Since the critical concentration $c^*$ has been well investigated \cite{Tirtaatmadja2006},  the Mark–Houwink relation as intrinsic viscosity $[\eta]=0.072M_w^{0.65}$ was adopted, suggesting the critical concentration $c^*=0.047$ wt.\%, 0.195 wt.\% and 0.93 wt.\% for 8 M, 1 M and 0.1 M, respectively.
Hence, the corresponding concentration of PEO is 0.04 wt.\%, 0.19 wt.\%, and 0.93 wt.\%, respectively. The solution was re-stirred before each measurement and used within two weeks after preparation to avoid degradation effects.

{PIB-PB Solution Preparation:}  We first prepared a 3 wt.\% polyisobutylene (PIB) solution by melting and dissolving the PIB polymer (average $M_w=4.2\times10^6$, Sigma-Aldrich) in light mineral oil (Sigma-Aldrich) at 100°C with heating and stirring. Next, we mixed polybutene (PB, average $M_n=2300$, Sigma-Aldrich) and light mineral oil as the solvent, to which the previously prepared PIB-mineral oil solution was added. The final solution, containing {0.4, 0.3, 0.2, 0.02} wt.\% PIB polymer with 60 wt.\% PB and 40 wt.\% light mineral oil as the solvent, was obtained through mechanical stirring followed by vacuum degassing.

{PS-DOP Solution Preparation:} Polystyrene (PS, average $M_w = 2 \times 10^6 \, \mathrm{g/mol}$, Sigma-Aldrich) was dissolved in dioctyl phthalate (DOP, Sigma-Aldrich) at a mass fraction of 0.1\%. The mixture was stirred until fully dissolved. Notably, PS-DOP is known to be a theta system at $22^{\circ}$C \cite{johnson1970}. The solution was further diluted stepwise by adding DOP to prepare samples with mass fractions of 0.05 wt.\% and 0.025 wt.\%. 

\begin{figure*}
\includegraphics[width=\textwidth]{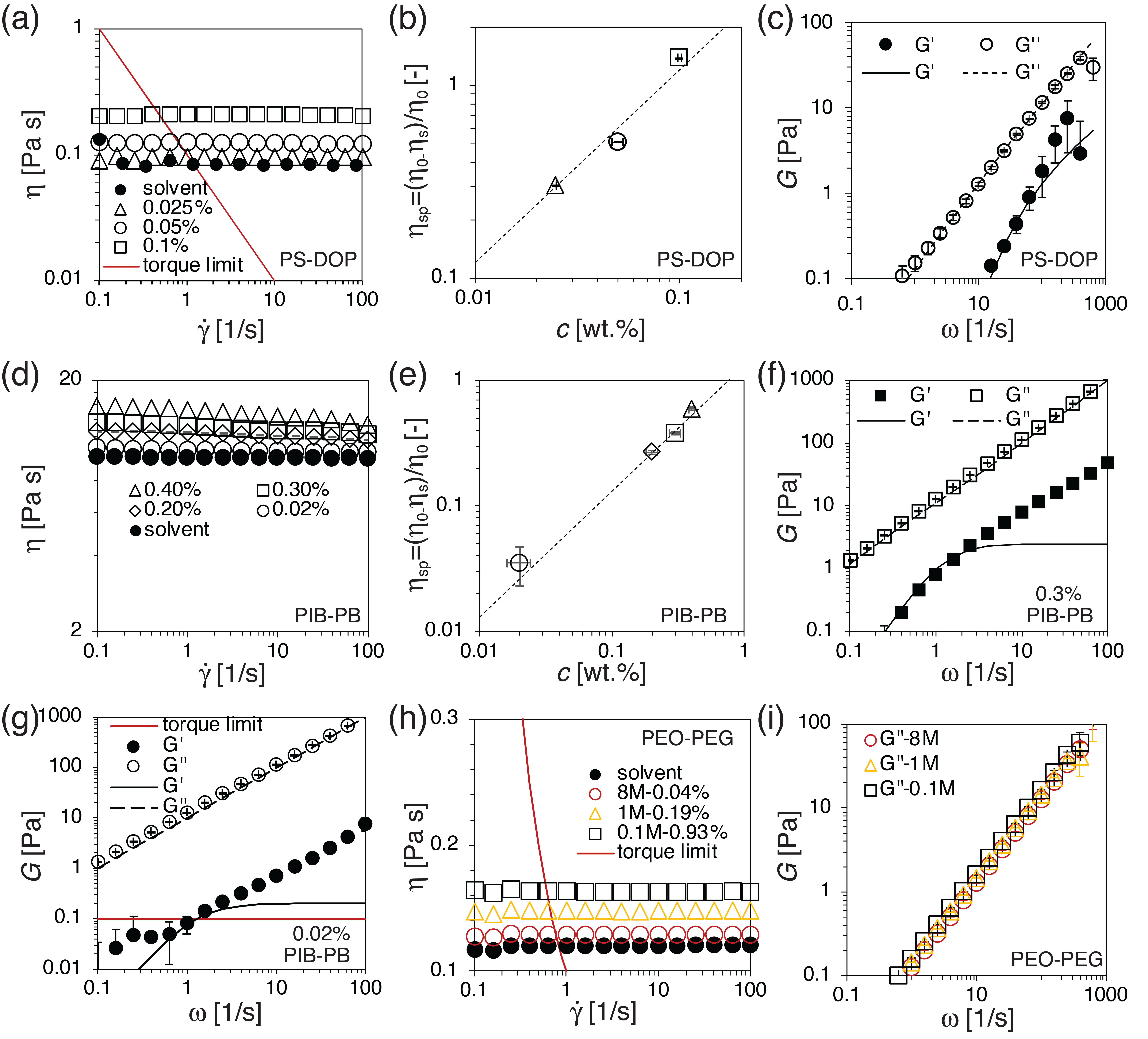}
\caption{\label{fig:FigureS2}  
Rheological characterization of the test fluids, including PS-DOP solutions (a–c), PIB-PB solutions (d–g), and PEO-PEG solutions (h–i). 
(a) Viscosity $\eta$ as a function of shear rate $\dot{\gamma}$ at various concentrations for PS-DOP (a), PIB-PB (d), and PEO-PEG (h) solutions. 
(b) Specific viscosity, defined as $\eta_{sp} = (\eta_0 - \eta_s)/\eta_s$, versus concentration $c$, where a power-law exponent of 1 indicates the dilute regime. 
(c) Measurement of the loss modulus ($G''$) and storage modulus ($G'$) under small-amplitude oscillatory shear for PS-DOP at $c = 0.05$ wt.\% (c), PIB-PB at $c = 0.3$ wt.\% (f) and $c = 0.02$ wt.\%, and PEO-PEG with various molecular weights of PEO (i). 
In all figures, the absence of error bars indicates that the error bars are smaller than the symbol size. 
The curves in (c) are fitted using the Zimm model, while those in (f, e) are fitted using the Oldroyd-B model.
}
\label{fig:S2}
\end{figure*}

The viscosity and modulus for each  solution was measured using a rheometer (Anton Paar 302e). For viscosity testing, a cone-and-plate configuration was used (cone angle $\alpha=1^{\circ}$, plate radius $R = 25 \, \mathrm{mm}$) for PS-DOP and PIB-PB solutions, while a double-gap configuration was used (DG26.7) for PEO-PEG solutions so as to prevent  evaporation effects.
The results, shown in Fig.~\ref{fig:S2}(a,d,h), indicate that the viscosity of solutions remains nearly constant with varying shear rates, suggesting that they can be considered as Boger fluids. 
Measured specific viscosity $\eta_{sp}=(\eta_0-\eta_s)/\eta_0$ validated that all PS-DOP (Fig.~\ref{fig:S2}b) and PIB-PB solutions (Fig.~\ref{fig:S2}e) remain within the dilute regime due to $\eta_{sp}\sim c$. Here we choose and label 0.05 wt.\% PS solution as PS-DOP, 0.3 wt.\% PIB solution as PIB-PB-0.3, and 0.02 wt.\% PIB solution as PIB-PB-0.02 for the test fluids.

 Using the same configuration settings, small-amplitude oscillatory rheology tests were conducted to measure the loss modulus ($G''$) and storage modulus ($G'$). The results are shown in Fig.~\ref{fig:S2}(c,f,g,i), corresponding to the PS-DOP, PIB-PB-0.3, PIB-PB-0.02, and PEO-PEG solutions, respectively. In Fig.~\ref{fig:S2}(c), the curves were fitted using the Zimm model, while in Fig.~\ref{fig:S2}(f,g), the curves were fitted using the Oldroyd-B model. The formulas for the moduli under small-amplitude oscillation for the Oldroyd-B model are:
\begin{subequations}
    \begin{equation}
G^{\prime}(\omega)= \frac{G_p \lambda_s^2 \omega^2}{1+\left(\lambda_s \omega\right)^2}
\end{equation}
\begin{equation}
G^{\prime\prime}(\omega)= \eta_s\omega+\frac{G_p\lambda_s  \omega}{1+\left(\lambda_s \omega\right)^2}.
\end{equation}
\end{subequations}
For the Zimm model the equations are:
\begin{subequations}
    \begin{equation}
G^{\prime}\left(\omega\right)=G_p \frac{\omega \lambda_s \sin \left[\chi \operatorname{atan}\left(\omega \lambda_s\right)\right]}{\left[\left(1+\left(\omega \lambda_s\right)^2\right)\right]^{\chi / 2}},
\end{equation}
\begin{equation}
G^{\prime\prime}\left(\omega\right)=\eta_s\omega+G_p\frac{\omega\lambda_s \cos \left[\chi \operatorname{atan}\left(\omega \lambda_s\right)\right]}{\left[\left(1+\left(\omega \lambda_s\right)^2\right)\right]^{\chi / 2}},
\end{equation}
\end{subequations}
where $\omega$ is the frequency, $\lambda_s$ is the relaxation time for small-amplitude-oscillatory-shear rheometry, $G_p=\eta_p/\lambda$, and $\chi=1/3$ at the theta temperature \cite{rubinstein2003}.

The surface tension of all studied  liquids was measured using the pendant drop method based on the initial droplet shape in the dripping tests.
The physical property data for all tested fluids are summarized in the Table~\ref{tab:fluid_properties}.
\begin{table*}
\caption{\label{tab:fluid_properties}%
Summary of physical property data for the fluids used in this study. Viscosity $\eta_0$ and solvent viscosity $\eta_s$, viscosity ratio $\beta=\eta_s/\eta_0$, and surface tension $\gamma$ are reported. Fitting relaxation time $\lambda_s$ based on Oldroyd-B model or Zimm model from small amplitude oscillatory shear rheometry.}
\begin{ruledtabular}
\begin{tabular}{lccccc}
\textbf{Name} & 
{$\eta_0$ [Pa s]} & 
{$\eta_s$ [Pa s]} & 
{$\beta$} & 
{$\gamma$ [mN/m]} & $\lambda_s$ [s] \\
\hline
PIB-PB-0.3& $12.88 \pm 0.08$ & $9.95 \pm 0.04
$ & $0.773\pm0.006$ 
& $28.56 \pm 1.18$ & $0.7\pm0.1$ \\
PIB-PB-0.02& $10.22 \pm 0.01$ & $9.95 \pm 0.04
$ & $0.974\pm0.004$ 
& $29.55 \pm 1.62$ & $0.7\pm0.1$ \\
PS-DOP & $0.124 \pm 0.001$ & $ 0.081\pm0.001$
& $0.653$$\pm0.010$ 
& $31.78 \pm 3.14$ & $0.030\pm0.002$ \\
PEO-PEG-8M & $0.128 \pm 0.001$ & $0.127 \pm 0.001$ & $0.985$ $\pm 0.011$ 
& $50.26 \pm 2.10$ & / \\
PEO-PEG-1M & $0.148 \pm 0.001$ & $0.127 \pm 0.001$ & $0.858\pm 0.009$ 
& $46.62 \pm 1.50$ & / \\
\end{tabular}
\end{ruledtabular}
\end{table*}

\section{III. Experimental Setup and Procedure}
\begin{figure}
\includegraphics[width=\columnwidth]{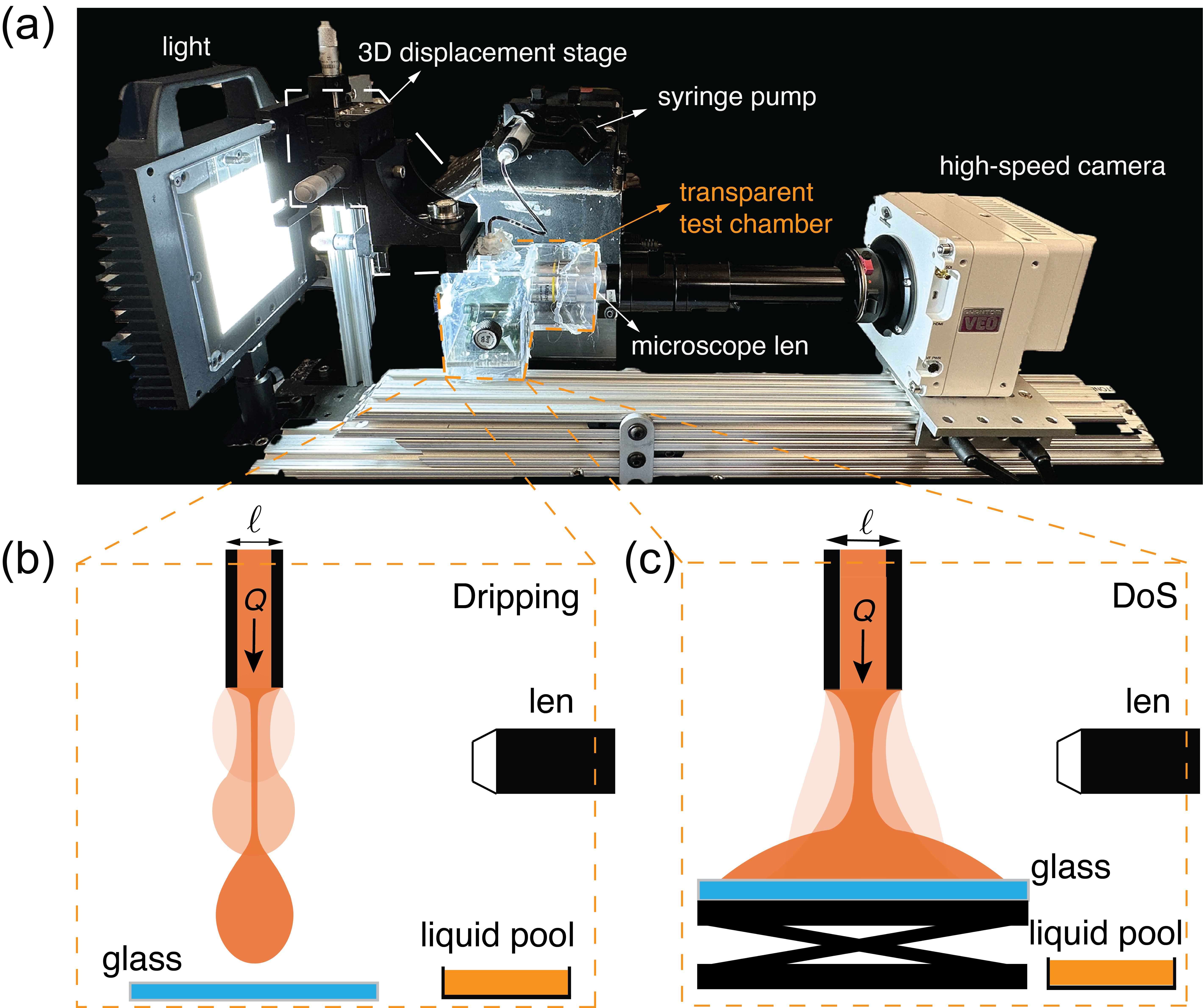}
\caption{\label{fig:FigureS3}  
Experimental setup and schematic diagrams of dripping and DoS methods. (a) The capillary breakup testing apparatus consists of a high-speed camera equipped with a microscope objective, a syringe pump, a three-dimensional displacement stage, and a light source. A transparent testing chamber is used to prevent environmental airflow disturbances and to reduce evaporation of the test liquid. (b) Schematic of the dripping method. (c) Schematic of the DoS method, where the substrate is made of quartz glass. A liquid reservoir inside the chamber is included to create a saturated vapor atmosphere.
}
\label{fig:S3}
\end{figure}

The experimental setup for capillary breakup tests used in this study is shown in Fig.~\ref{fig:S3}(a). The system consists of a high-speed camera (640L, Phantom), microscope objectives (2X, 5X, and 10X, Nikon, to accommodate different needle diameters), a syringe pump (Harvard 2000), an LED backlight source (130,000 LUX, PHLOX), and a custom-made transparent chamber. As illustrated in the schematics shown in Figs.~\ref{fig:S3}(b) and (c), both the needle and microscope objective are positioned inside the transparent chamber for the dripping and DoS tests. The sealed chamber prevents environmental airflow from disturbing the filament during capillary breakup. Additionally, a liquid reservoir containing the test fluid is placed inside the chamber. Before each test, the chamber is allowed to equilibrate for several hours, ensuring the space is saturated with the solvent vapor of the test liquid, thereby minimizing the effects of evaporation during the experiment.

\begin{figure*}
\includegraphics[width=\textwidth]{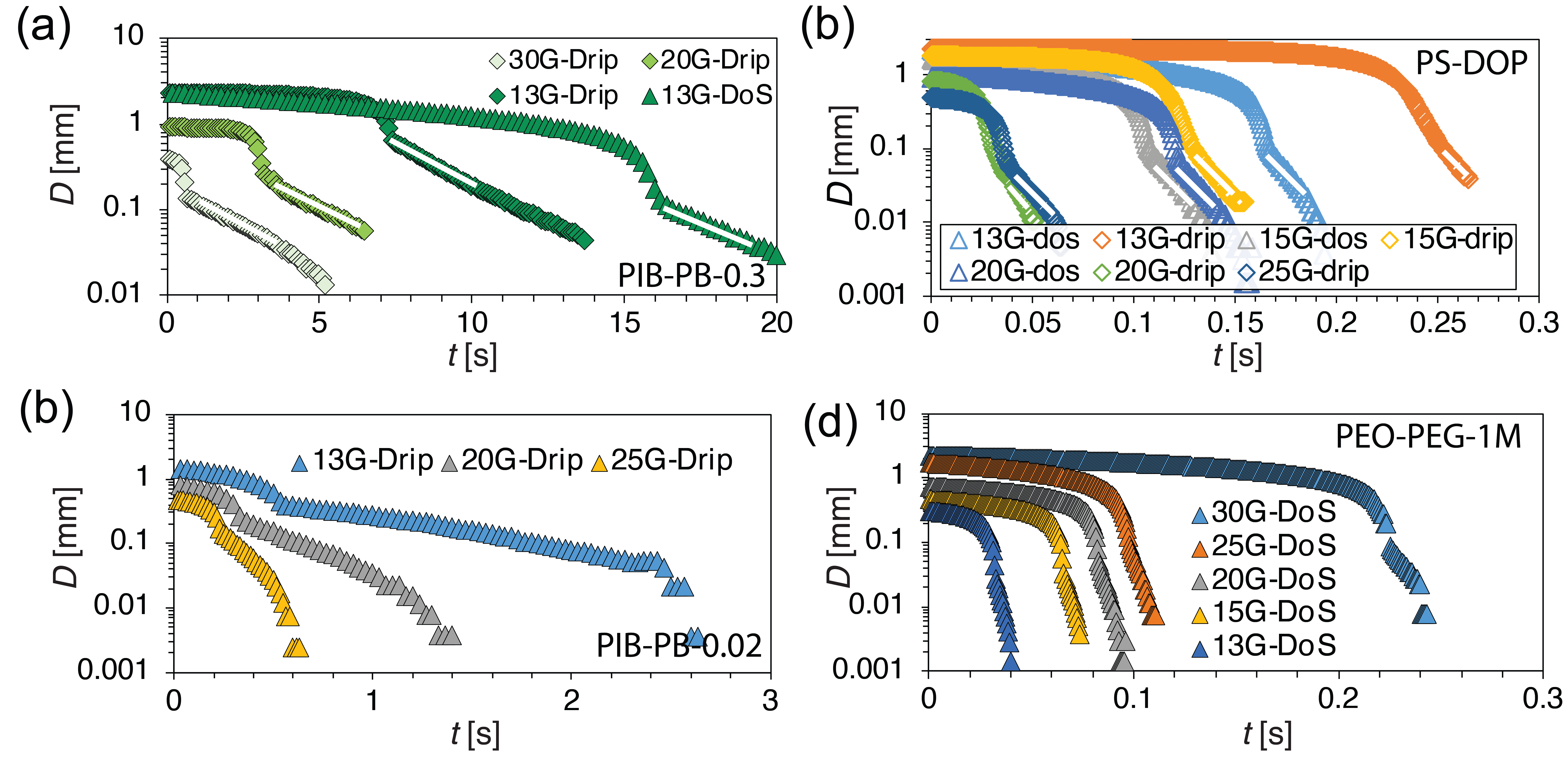}
\caption{\label{fig:FigureS4}  
Representative experimental results of filament diameter evolution over time for 
(a) PIB-PB-0.3, (b) PS-DOP, (c) PIB-PB-0.02, and (d) PEO-PEG-1M solutions.
}
\label{fig:S4}
\end{figure*}

Standard syringe needles were used as nozzles in the experiments, with models including \{30G, 25G, 20G, 15G, 13G\}, corresponding to outer diameters $\ell \in \{0.305, 0.508, 0.902, 1.829, 2.413\} \, \mathrm{mm}$. The controlled flow rate during the experiments was set to $Q = 1-10 \, \mu\mathrm{L/min}$ to ensure that the hanging droplet extruded from the nozzle remained quasi-static, avoiding pre-stretching of the polymer induced by injection.
The images capturing the liquid thinning and breakup processes were batch-processed using a custom MATLAB program. This program was used to track the liquid interface and extract the temporal evolution of the minimum diameter $D(t)$. Representative results from the dripping and DoS tests on PEO-PEG, PIB-PB, and PS-DOP solutions are presented in Fig.~1 of the main text and Figs.~\ref{fig:S4}. Each test was repeated three times to ensure the reproducibility of the $D(t)$ data. 
\section{IV. Analytical Model and Numerical Results}

When the droplet or liquid bridge is stretched beyond the critical condition for the  Rayleigh-Plateau instability, the break-up process will be dominated by viscous, capillary, and inertial forces for Newtonian fluids, which can be categorized as visco-capillary (VC) or inertial-capillary (IC) regimes depending on the Ohnesorge number ($Oh \equiv \eta/\sqrt{\rho \mathcal{R} \gamma}$) according to $Oh \gg 1$ or $Oh \ll 1$ respectively, where $\eta$, $\rho$, $\gamma$ represent viscosity, density, and surface tension. $\mathcal{R}$ is the characteristic size of the droplet or liquid bridge. 

The physical principle of capillary breakup rheometry for a viscoelastic fluid is based on the elasto-capillary (EC) dynamics of the liquid bridge/filament after IC or VC. There is a widely-used formula in numerous papers to calculate relaxation time $\lambda$ as
\begin{equation}
    \frac{R(t)}{R_1} = \left(\frac{G_p R_1}{2\gamma}\right)^{1/3}\exp{\left[\frac{-(t-t_1)}{3\lambda}\right]},
    \label{EC}
\end{equation}
where $R(t)$ is the minimum radius of filament, $G_p\equiv \eta_p/\lambda$ modulus of viscoelastic fluids, $t_1$ is the onset of the EC regime, and $R_1$ is the radius at this time. However, there are some intrinsic problems about this formula as given in the following discussion.

\subsection{A. Zero-dimensional Model }
Consider a cylindrical liquid filament with radius $R$ and length $L$ in the axisymmetric coordinate of $(r,z)$. The time-dependent relationship between $R$ and $L$ in cylinder deformation is 
\begin{equation}
    \pi R^2 L = \pi (R+\dot R \Delta t)^2(L+\dot L \Delta t) \Rightarrow \dot \epsilon \equiv \frac{\dot L}{L} = -\frac{2}{R} \dot R,
\end{equation}
where $\dot R \equiv dR/dt$, $\dot L \equiv dL/dt$, $\Delta t$ is a differential time, and $\dot \epsilon$ is the extensional strain rate. 
The velocity field in the filament thinning flow can be considered as $(v_r,v_z)=(-r\dot\epsilon/2,  z\dot\epsilon)$. Hence, 
the momentum equation neglecting inertial effects for $Oh\gg1$ is
\begin{subequations}
    \begin{equation}
        -\frac{\gamma}{R} = \sigma_{rr} = -p - \eta_s \dot \epsilon + \sigma_{p,rr}
    \end{equation}
    \begin{equation}
        0 = \sigma_{zz} = -p + 2\eta_s \dot \epsilon + \sigma_{p,zz},
    \end{equation}
\end{subequations}
where ${\sigma}$ is the stress and subscript $p$ is the polymeric part. The combination of the above two equations leads to 
\begin{equation}
    \frac{\gamma}{R} = 3 \eta_s \dot \epsilon -\sigma_{p,rr}+\sigma_{p,zz}.
    \label{total_force_balance}
\end{equation}
Based on the FENE-P model, the polymeric stress $\sigma_p$ can be written as \cite{Alves2021}
\begin{equation}
     \sigma_p =  \frac{\eta_p}{\lambda} \left( f \boldsymbol{A} - \boldsymbol{I} \right) = \frac{\eta_p}{\lambda} \left( \frac{b-3}{b - \mathrm{tr}(\boldsymbol{A})} \boldsymbol{A} - \boldsymbol{I} \right)
     \label{polymericStressFENEP}
\end{equation}
where $\mathrm{tr}(\boldsymbol{A})$ denotes the trace of the conformation tensor $\boldsymbol{A}$ and $b$ is the extensibility parameter. If $f=1$, Eq. (\ref{polymericStressFENEP})  will reduce to the Oldroyd-B model, i.e. $b=\infty$. The conformation tensor $\boldsymbol{A}$ obeys
\begin{equation}
\overset{\bigtriangledown}{\boldsymbol{A}} = -\frac{1}{\lambda} \left( f \boldsymbol{A} - \boldsymbol{I} \right),
\label{conformantino}
\end{equation}
where $\overset{\bigtriangledown}{\boldsymbol{A}}\equiv \partial_t \boldsymbol{A} + (\boldsymbol{v} \cdot \boldsymbol{\nabla})\boldsymbol{A}-(\boldsymbol{\nabla v})\cdot\boldsymbol{A} - \boldsymbol{A}\cdot(\boldsymbol{\nabla v})^T$ is the upper convected derivative of the conformation tensor.
Substitution of the flow field ($v_r$, $v_z$) into (\ref{conformantino}) and combining with the momentum equation yields
\begin{subequations}
    \begin{equation}
\frac{\gamma}{R}=3 \eta_s \dot{\epsilon}+\frac{\eta_p}{\lambda}f\left(A_{zz}-A_{rr}\right)
\label{A_force}
\end{equation}
\begin{equation}
\lambda \dot A_{z z}=(2 \dot{\epsilon} \lambda-f) A_{z z}+1
\label{A_zz_time}
\end{equation}
\begin{equation}
\lambda\dot A_{r r}=-(\dot{\epsilon} \lambda+f) A_{r r}+1 
\end{equation}
\begin{equation}
    f =\frac{b-3}{b-A_{zz}-2A_{rr}}
\end{equation}
\begin{equation}
    \dot \epsilon =-2 \frac{\dot R}{R}.
\end{equation}
\label{original_GE}
\end{subequations}
This set of equations describes the variations of radius $R(t)$, extensional strain rate $\dot\epsilon(t)$, and conformation components $A_{zz}(t)$ and $A_{rr}(t)$ as a function of time $t$.

\subsection{B. Simplified Analytical Solutions}
\subsubsection{1. Infinite extensibility length, i.e., $b=\infty$}
Adopting assumptions of (i) $f=1$ for infinite $b$ and (ii) $A_{zz} \gg 1$, we can obtain an ODE for $A_{zz}$ 
\begin{equation}
    \dot A_{zz} = 2\dot\epsilon A_{zz}-\frac{1}{\lambda}A_{zz} = -\frac{4\dot R}{R}A_{zz}-\frac{1}{\lambda}A_{zz},
\end{equation}for 
which  the time-dependent solution  $A_{zz}(t)$ with the initial conditions $A_{zz}(t_1)=1$ and $R(t_1)=R_1$ is
\begin{equation}
    A_{zz} = \frac{R_1^{4}}{R^4(t)}\exp{\left(-\frac{t-t_1}{\lambda}\right)}.
    \label{A_zz}
\end{equation}
Substituting (\ref{A_zz}) into (\ref{A_force}) with {two other assumptions  of  (iii) negligible viscous effects,  $\sigma_s\equiv3\eta_s\dot\epsilon \ll \sigma_{p,zz}$, and (iv) $A_{zz} \gg A_{rr}$ (implying $\sigma_{p,zz} \gg \sigma_{p,rr}$)}, the final formula for $R(t)$ is
\begin{equation}
    \frac{R(t)}{R_1}
    =\left(\frac{\eta_p R_1}{\lambda \gamma}\right)^{1/3}\exp{\left(\frac{t_1-t}{3\lambda}\right)}.
\end{equation}
Hence, the extensional strain rate $\dot \epsilon$ and $A_{zz}$ during EC is
\begin{equation}
    \dot \epsilon = -2 \dot R/R = \frac{2}{3\lambda}
\end{equation}
\begin{equation}
    A_{zz} = \left(\frac{G R_1}{\gamma}\right)^{-4/3}\exp{\left(\frac{t-t_1}{3\lambda}\right)}.
\end{equation}
This formula was first derived by Entov and Hinch \cite{Entov_1997} in 1997.
Defining the Deborah number $De \equiv \dot \epsilon \lambda$, we can find a constant $De = 2/3$ in the EC regime as derived as 
\begin{equation}
    De = \dot \epsilon \lambda = -\frac{2\dot R}{R}\lambda = \frac{2}{3} .
\end{equation}
However, it is well-known that the Oldroyd-B model fails to provide the correct prediction of polymer extension for $De > 1/2$ (the coil-stretch transition). Thus, there is inconsistency.

\subsubsection{2. Finite extensibility $b$ with unknown $A_{zz}(t_1)$ }
\begin{figure}
    \centering
    \includegraphics[width=\linewidth]{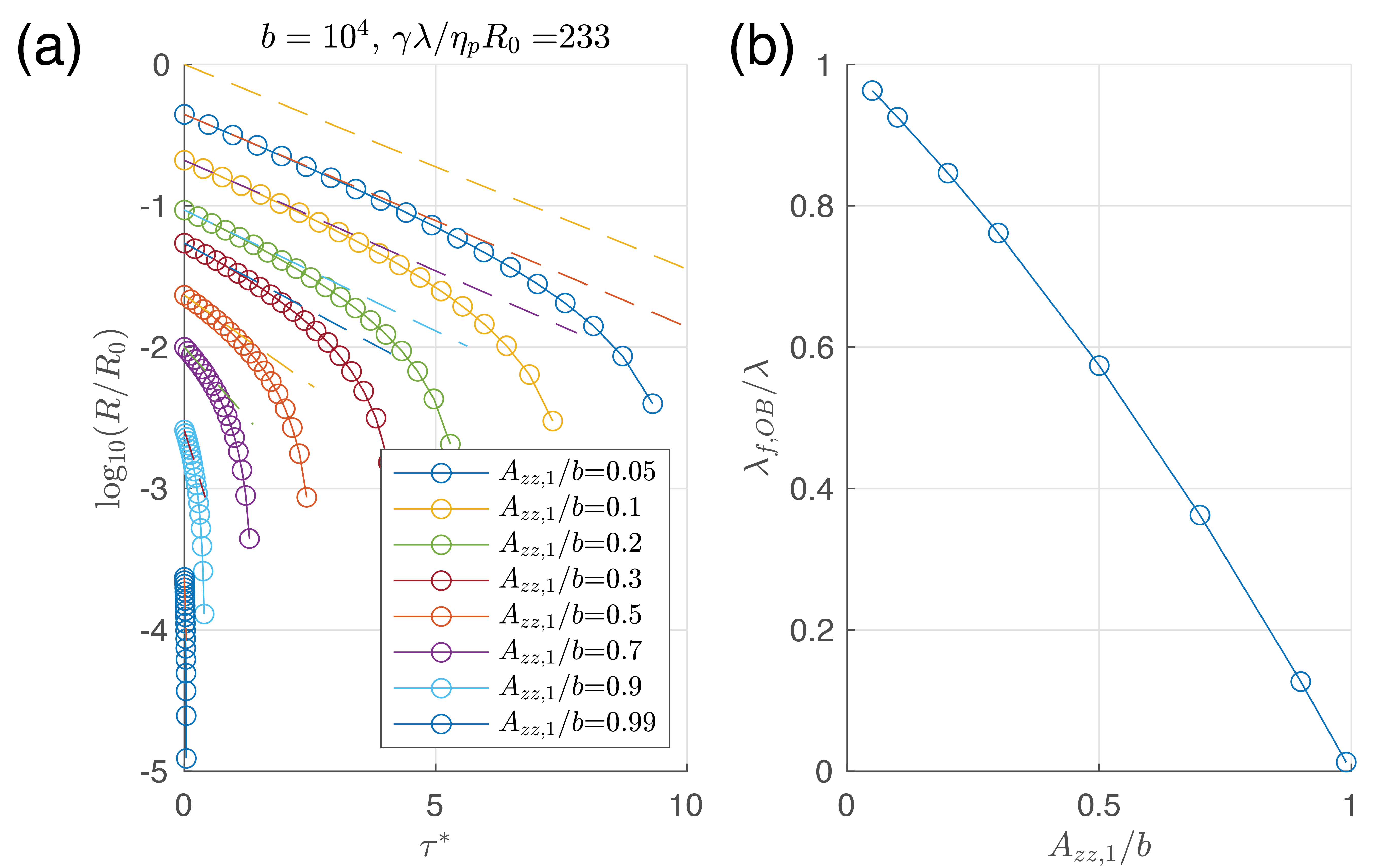}
    \caption{(a) Filament radius $R$ evolves at various pre-stretched conditions of $A_{zz,1}$ and (b) corresponding fitting $\lambda_{f,OB}$}.
    \label{FigureS5}
\end{figure}

With the same assumptions (ii) $A_{zz} \gg 1 > A_{rr}$ and (iii) $\sigma_p\gg\sigma_s$,  and assuming (v) $b$ is finite and $b\gg3$ for the EC regime, we can derive 
\begin{equation}
    \frac{\gamma}{R} = \frac{\eta_p}{\lambda} fA_{zz} \Rightarrow R = \frac{\gamma \lambda}{\eta_p} \left( \frac{1}{A_{zz}} - \frac{1}{b} \right).
    \label{total_balance_FENE_P}
\end{equation}
Therefore, we obtain
\begin{equation}
    \dot R = -\frac{\gamma \lambda}{\eta_p} \frac{\dot A_{zz}}{A_{zz}^2} \Rightarrow \dot \epsilon = \frac{2 \dot A_{zz}}{A_{zz} \left( 1 - A_{zz}/b \right)}.
    \label{dot_R}
\end{equation}
Substituting (\ref{dot_R}) into (\ref{A_zz_time}) yields 
\begin{equation}
    \frac{A_{zz}}{\lambda} = \dot A_{zz} \left( 3 + \frac{A_{zz}}{b} \right),
\end{equation}
which has an implicit solution, also as reported in \cite{Gaillard_2024}, given by
\begin{equation}
    \frac{t - t_1}{\lambda} = \frac{A_{zz} - A_{zz,1}}{b} + 3 \ln{\left( \frac{A_{zz}}{A_{zz,1}} \right)},
\end{equation}
where the subscript 1 represents the onset of EC, and $A_{zz,1}$ denotes the stretching state at that time. 
Recalling (\ref{total_balance_FENE_P}) and taking the logarithm of both sides of the equation, we obtain
\begin{equation}
        \log_{10}{\left(R/R_{1}\right)}= \log_{10}{\left(\frac{1}{A_{zz}}-\frac{1}{b}\right)}+ \log_{10}{\left(\frac{\gamma\lambda}{\eta_pR_{1}}\right)}
        \label{R_R_1_fene_t}
\end{equation}
The numerical results of (\ref{R_R_1_fene_t}) are shown in Figure \ref{FigureS5}a. By taking the tangent line at the initial point for fitting based on the Oldroyd-B model, it indicates that the fitting relaxation time $\lambda_{f,OB}$ is always less than the true relaxation time $\lambda$ and decreases with increasing $A_{zz,1}$ (Figure \ref{FigureS5}b).

Although this implicit expression captures the reduction of the apparent relaxation time due to finite extensibility $b$, the inability to determine the initial value $A_{zz,1}$ in the EC phase introduces a new fitting parameter $A_{zz,1}$ if this expression is used to interpret experimental data.

\subsubsection{3. Piecewise Approximation Considering Pre-Stretching}
One approach to account for the pre-stretching process is to treat the VC and EC stages separately, i.e. ``piecewise approximation'', using the endpoint of the VC stage as the initial condition for the EC stage. A natural way to implement this is to determine the transition point based on the relative magnitudes of the polymeric stress $\sigma_p$ and the viscous stress $\sigma_s$ \cite{Wagner2015}. Below, we demonstrate how to derive the corresponding approximate formula based on this method.
If $3\eta_s \dot\epsilon \gg \sigma_{p,zz} - \sigma_{p,rr}$, the Eq.~(\ref{A_force}) simplifies to:
\begin{equation}
    \gamma/R = 3\eta_s \dot\epsilon  \Rightarrow R = (t_{c}-t)\gamma/6\eta_s,
\end{equation}
where $t_{c}=6\eta_sR_0/\gamma$. 
Although the elastic force has a minor influence during the VC regime, the polymer stretches along with the transient extensional strain rate, $\dot \epsilon = 2/(t_{c} - t)$. 
\begin{figure}
    \centering
    \includegraphics[width=\linewidth]{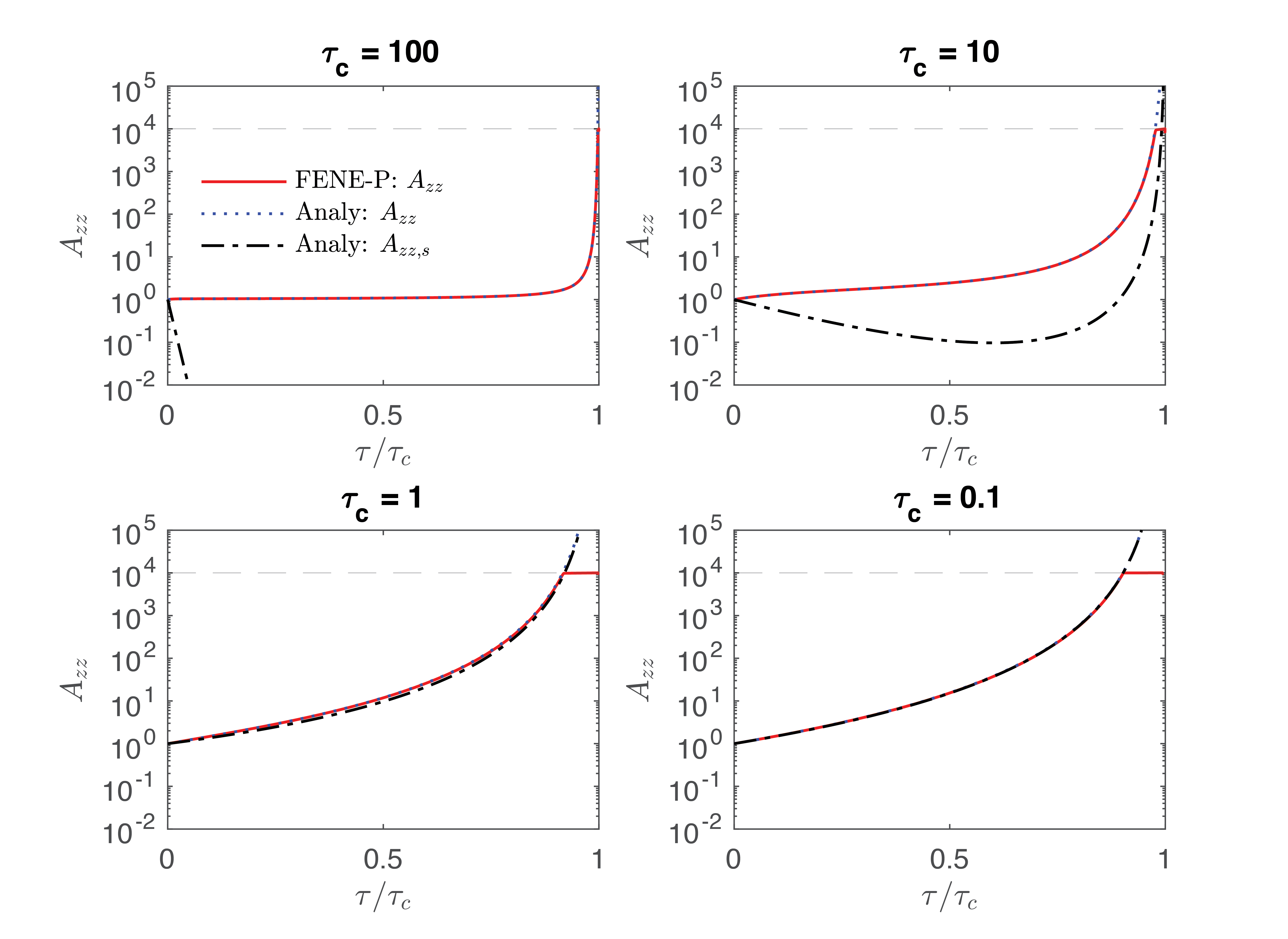}
    \caption{Transient components of $\boldsymbol{A}(t)$ in IC or VC regime for various $\tau_c$ when $\beta=0.9$ and $b=10^4$}.
    \label{FigureS6}
\end{figure}
Then substituting $\dot \epsilon = 2/(t_{c}-t)$ into (\ref{A_zz_time}) yields
\begin{subequations}
    \begin{equation}
        \partial_\tau A_{rr} = -\left(\frac{2}{\tau_c-\tau}+f\right)A_{rr}+1
    \end{equation}
    \begin{equation}
        \partial_\tau A_{zz} = \left(\frac{4}{\tau_c-\tau}-f\right)A_{zz}+1,
    \end{equation}
    \label{VC_comformation}
\end{subequations}
where $\tau = t/\lambda$ and $\tau_c=t_c/\lambda=6\eta_s R_0/\gamma\lambda$ in the EC regime, and $f=(b-3)/(b-\mathrm{tr}\,\boldsymbol{A})$ for the FENE-P model or $f=1$ for the Oldroyd-B model. This set of equations cannot be solved analytically  for  $\mathrm{tr}\boldsymbol{A}$.

A typical estimate gives $\mathcal{O}(\tau_c) \in [0.01,1]$ for dilute water-based polymer solutions, where $\eta_s \in [10^{-2},10^{-1}]$ Pa$\cdot$s, $\gamma \in [30,70]$ mN$\cdot$m, $\lambda \in [10,100]$ ms, and $R_0 \in [0.3,3]$ mm. This set of equations can be solved numerically with the initial condition $\boldsymbol{A}=\boldsymbol{I}$. Furthermore, we can estimate $A_{zz}(t)$ using the Oldroyd-B model. The analytical solution for $f=1$ is
\begin{subequations}
    \begin{equation}
        \begin{aligned}
        A_{zz}(\tau) = & \frac{e^{-\tau}}{(\tau_c -\tau)^4} \left[-24 + e^\tau \left(\tau_c^4 + 24 - 24\tau + 12\tau^2 - 4\tau^3 + \tau^4\right) \right. \\
        & - 4\tau_c^3 \left(1 + e^\tau (\tau - 1)\right) + 6\tau_c^2 \left(-2 + e^\tau (2 - 2\tau + \tau^2)\right) \\
        & \left. - 4\tau_c \left(6 + e^\tau (-6 + 6\tau - 3\tau^2 + \tau^3)\right) \right]
        \end{aligned}
        \label{A_zz_OB}
    \end{equation}
    \begin{equation}
        \begin{aligned}
    A_{rr}(\tau) &= \frac{e^{-\tau} (\tau_c - \tau)}{\tau_c^2} \Bigg[ (\tau_c - \tau) + \tau_c^2 + \tau_c^2 e^\tau - \tau + \tau_c \tau \\
    &\quad - \tau_c^2 e^{\tau_c} (\tau_c - \tau) \text{Ei}(-\tau_c)  
    \\
    &\quad
    + \tau_c^2 e^{\tau_c} (\tau_c - \tau) \text{Ei}(-\tau_c + \tau) \Bigg],
        \end{aligned}
    \end{equation}
\end{subequations}
where the integral function $\operatorname{Ei}(x) = -\int_{-x}^{\infty} \frac{e^{-t}}{t} \, \mathrm{d}t$. If $4/\tau_c \gg 1$ so that  $f$ may be neglected in the Eq.(\ref{VC_comformation}), these equations can be further simplified as
\begin{subequations}
    \begin{equation}
    A_{zz,s}(\tau) = \frac{\tau_c^4 \exp(-\tau)}{\left(\tau_c-\tau\right)^4}
    \label{A_zz_OB_simple}
    \end{equation}
    \begin{equation}
    A_{rr,s}(\tau) = A_{\theta\theta,s}(\tau) = \frac{\left(\tau_c-\tau\right)^2 \exp(-\tau)}{\tau_c^2}.
    \end{equation}
\end{subequations}
Eq.(\ref{A_zz_OB_simple}) has been reported in \cite{Wagner2015}.
The numerical and analytical results of the $A_{zz}$ variation are displayed in Fig.~\ref{FigureS6}, which indicates that $A_{zz}$ predicted by the Oldroyd-B model can overlap with the FENE-P model before reaching a plateau.

\begin{figure}[h]
    \centering
    \includegraphics[width=\linewidth]{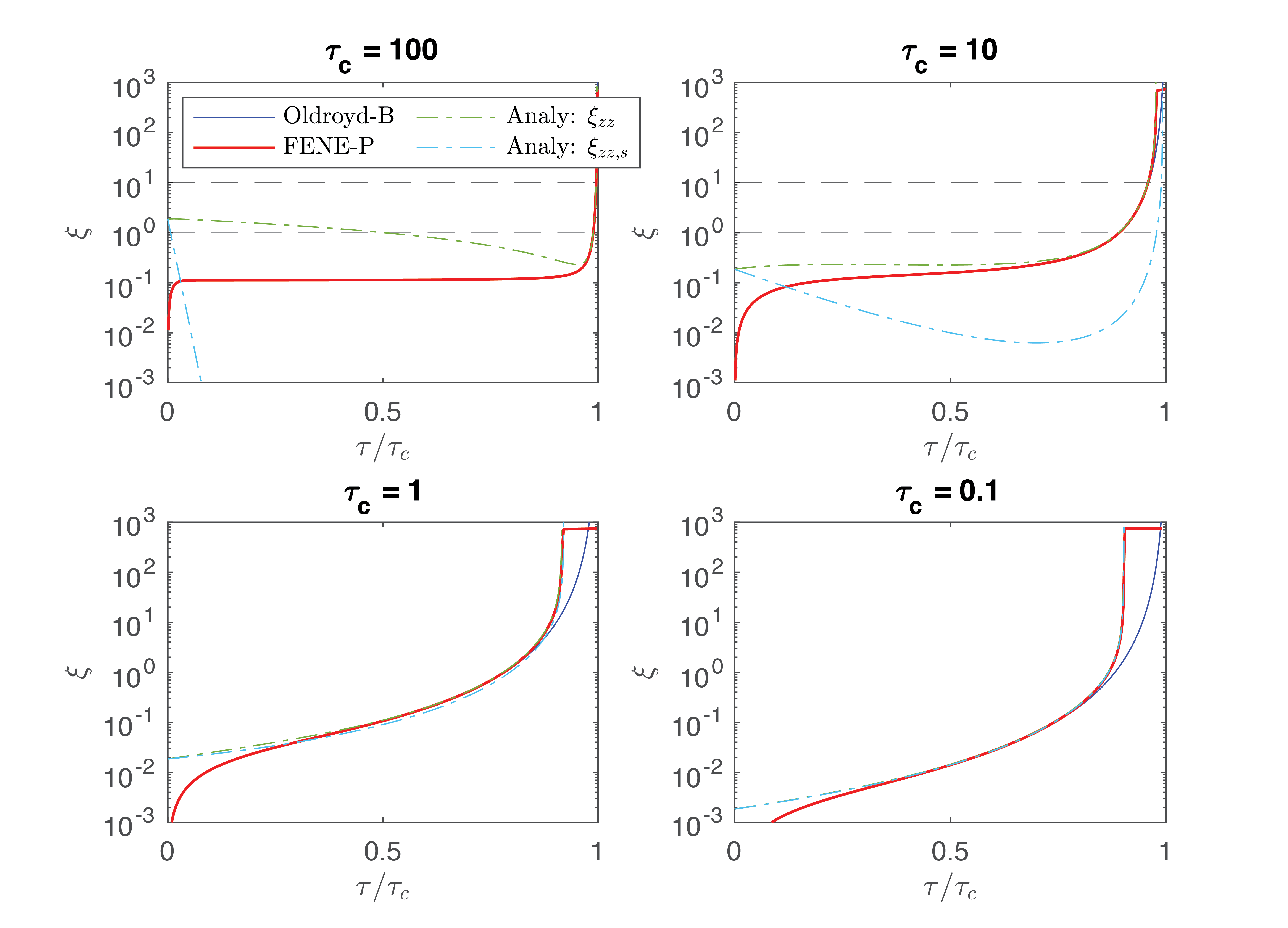}
    \caption{Ratio of transient polymeric stress $\sigma_{p,zz}-\sigma_{p,rr}$ to viscous stress $3\eta_s\dot\epsilon$ with the definition $\xi = (\sigma_{zz}-\sigma_{rr})/\sigma_s$}.
    \label{FigureS7}
\end{figure}

With the obtained $A_{rr}$ and $A_{zz}$, we can evaluate the ratio of transient polymeric stress $\sigma_{p,zz}-\sigma_{p,rr}$ to viscous stress $3\eta_s\dot\epsilon$ by denoting $\xi \equiv (\sigma_{zz}-\sigma_{rr})/\sigma_s = \eta_{p}(fA_{zz}-fA_{rr})/3\eta_s\dot\epsilon\lambda$, or
\begin{equation}
    \xi = \frac{1-\beta}{3\beta}f\frac{A_{zz}-A_{rr}}{\dot\epsilon(\tau)} = \frac{1-\beta}{6\beta}f(A_{zz}-A_{rr})(\tau_c-\tau).
\end{equation}
Considering $A_{rr} \ll A_{zz}$ in the late stage shown in Fig.~\ref{FigureS6}, we find the asymptotes $\xi_{zz}$ as 
\begin{subequations}
    \begin{equation}
        \begin{aligned}
        \xi_{zz} = & \frac{1-\beta}{6\beta}\frac{b-3}{b-A_{zz}}A_{zz}(\tau_c-\tau), \quad\text{for }  A_{zz} \gg A_{rr},
        \end{aligned}
        \label{xi_zz}
    \end{equation}
    \begin{equation}
    \xi_{zz,s} = \frac{1-\beta}{6\beta}\frac{\tau_c^4 \exp(-\tau)}{\left(\tau_c-\tau\right)^3}, \quad\text{for }  A_{zz} \gg A_{rr} \text{ and } \tau_c \ll 4.
\end{equation}
\end{subequations}
The numerical and analytical results of $\xi$ for $\beta=0.9$ and $b=10^4$ are depicted in Fig. \ref{FigureS7}.

Wagner et al. \cite{Wagner2015} suggested a threshold (denoted with *) by letting $\partial_{\tau^*} A_{zz}^*=0$, which leads to 
\begin{subequations}
    \begin{equation}
    0=2\dot\epsilon^*\lambda A_{zz}^*-(f^*A_{zz}^*-1) 
\end{equation}
\begin{equation}
    3\eta_s\frac{2}{\lambda(\tau_c-\tau^*)}=3\eta_s\dot\epsilon^*+\frac{\eta_p}{\lambda}f^*(A_{zz}^*-A_{rr}^*)
\end{equation}
\end{subequations}
Hence, the $\tau^*$ can be numerically determined from 
\begin{equation}
    \frac{A_{zz}^*}{b} = \frac{\frac{4}{\tau_c-\tau^*}-1}{\frac{4}{\tau_c-\tau^*}+\frac{2b\eta_p}{3\eta_s}}.
    \label{determine_tau_star}
\end{equation}
It should be emphasized that (\ref{A_zz_OB_simple}) was adopted to obtain the $\tau^*$ in (\ref{determine_tau_star}) instead of (\ref{A_zz_OB}) in the work of \cite{Wagner2015}, which is limited by the assumption of $\tau_c \ll 4$. Furthermore, the later results of the full model in Section \S IV-C show that the $\partial_{\tau^*} A_{zz}^*=0$ never occurs.

\begin{figure}
    \centering    \includegraphics[width=0.9\linewidth]{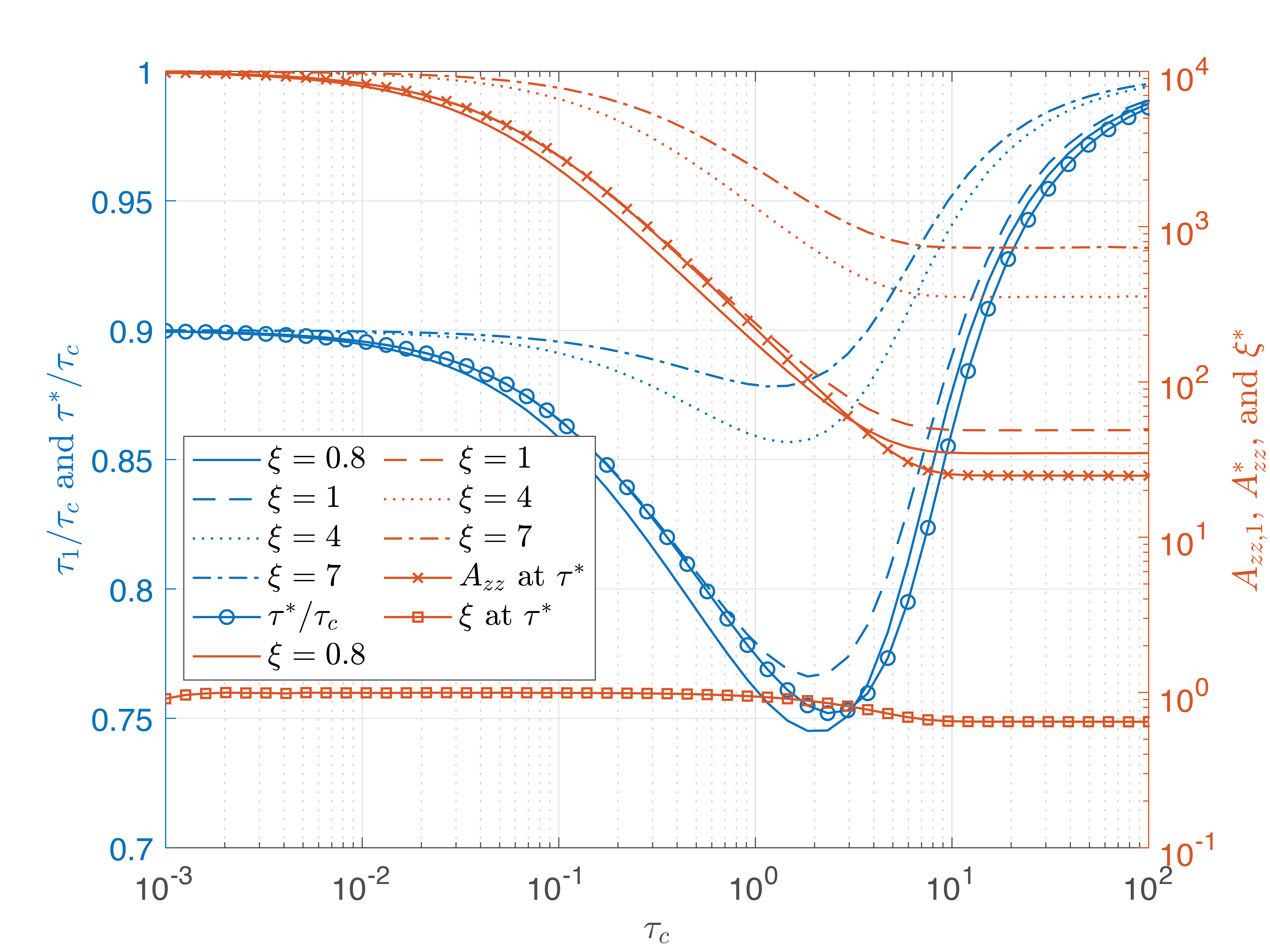}
    \caption{Relative transition time $\tau_{ec}/\tau_c$ and corresponding $A_{zz,1}$ for  various $\tau_c$ under different thresholds of $\xi$.}
    \label{FigureS8}
\end{figure}

In order to consider a range of $\tau_c$ from $10^{-3}$ to $10^2$, we here substitute (\ref{A_zz_OB}) into the above equation to numerically obtain $\tau^*$ and $A_{zz}^*$, which is also compared with results from cases of fixing $\xi$ to determine $\tau_{ec}$ and $A_{zz,1}$, as shown in Fig. \ref{FigureS8}. It can be concluded that $\partial_{\tau^*}A_{zz}^*$ leads to various $\xi \in (0.6, 1)$  for different $\tau_c$. The corresponding $A_{zz}^*$ is close to the case of $\xi \in (0.8, 1)$.

\begin{figure}
    \centering
\includegraphics[width=\linewidth]{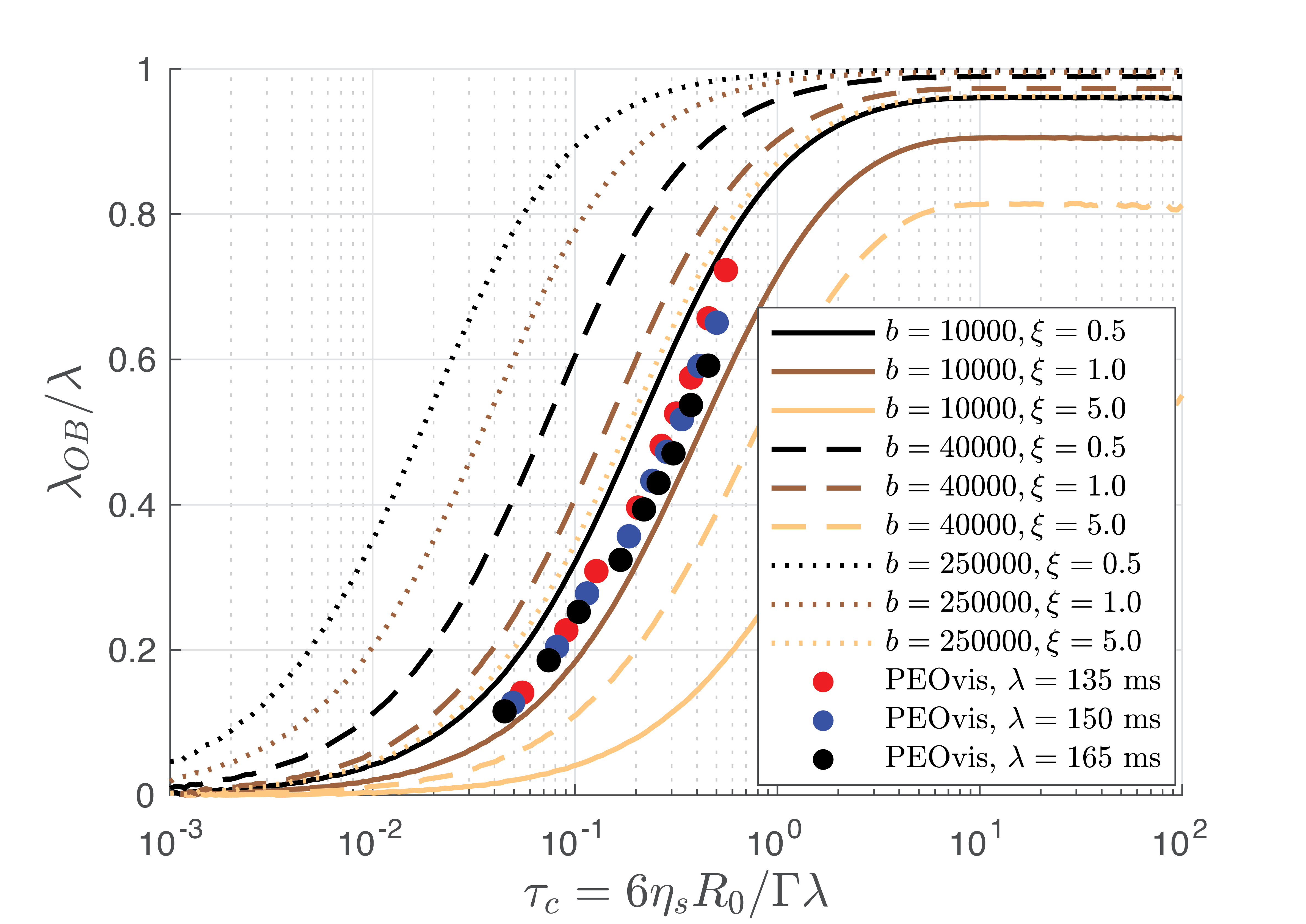}
    \caption{Relationship of $\lambda_{f,OB}/\lambda$ with $\tau_{c}$ for $b\in\{10^4,4\times10^4,2.5\times10^5\}$ and $\xi\in \{ 0.5,1,5\}$. Markers are experimental results of PEOvis \cite{Gaillard_2024} by assuming $\lambda\in\{135,150,165\}$ ms.}
    \label{FigureS9}
\end{figure}

Then we can input the results of $A_{zz,1}$ from Fig.  \ref{FigureS8} into (\ref{R_R_1_fene_t}) as initial conditions to obtain the relationship of $\lambda_{f,OB}/\lambda$ with $\tau_{c}$ as shown in Fig. \ref{FigureS9}.
Noting that for the pre-stretch stage, as described by the VC model, $\tau_c = 6\eta_sR_0/\gamma\lambda$, the results reveal the size effects of $R_0$, as displayed in Fig. \ref{FigureS9}.

However, this analysis introduces an additional variable, $\xi$, that must be considered. If the experimental results are interpreted using this approximate formula, as exemplified by PEOvis in Fig.~\ref{FigureS9}, the unknown parameters $\lambda$, $b$, and $\xi$ must be simultaneously accounted for. Under such conditions, determining the true relaxation time $\lambda$ becomes challenging.

\subsection{C. Full model}
By defining the dimensionless variables \(\tau = t/\lambda\), \(\beta = \eta_s/(\eta_p+\eta_s)\), \(a = R/R_0\), \(Ec = \lambda\gamma/\eta_pR_0\), and \(E = \dot\epsilon\lambda\), the original governing equations in (\ref{original_GE}) are transformed into their non-dimensional form as
\begin{subequations}
    \begin{equation}
    \frac{Ec}{a} = \frac{3\beta}{1-\beta}E + f(A_{zz}-A_{rr})
\end{equation}
\begin{equation}
    \partial_\tau A_{zz}   = \left(2E-f\right)A_{zz}+1
\end{equation}
\begin{equation}
  \partial_\tau A_{rr}   = \left(-E-f\right)A_{rr}+1
\end{equation}
\begin{equation}
    \partial_\tau a = -\frac{aE}{2}
\end{equation}
\label{dimensionless_eq}
\end{subequations}
where $f=(b-3)/(b-A_{zz}-2A_{rr})$, $b$ is the extensibility of the polymer.
\subsubsection{1. Analytical analysis}
Considering $b \sim A_{zz} \gg A_{rr}$ yields
\begin{subequations}
    \begin{equation}
        \frac{Ec}{a} = -\frac{6\beta}{1-\beta}\frac{1}{a}\frac{\mathrm{d}a}{\mathrm{d}\tau}+\frac{bA_{zz}}{b-A_{zz}}
    \end{equation}
    \begin{equation}
        \frac{\mathrm{d}A_{zz}}{\mathrm{d}\tau}=-\left(\frac{4}{a}\frac{\mathrm{d}a}{\mathrm{d}\tau}+\frac{1}{1-A_{zz}/b}\right)A_{zz}+1.
    \end{equation}
\end{subequations}
Letting $A_{zz}/b=\tilde{A}$ and $Ec/b = \tilde{Ec}$, this set of equations can be written as 
\begin{subequations}
    \begin{equation}
        \frac{6\beta}{1-\beta}\frac{1}{ab}\frac{\mathrm{d}a}{\mathrm{d}\tau}+\frac{\tilde{Ec}}{a}=\frac{\tilde{A}}{1-\tilde{A}}
        \label{a}
    \end{equation}
    \begin{equation}
        \frac{\mathrm{d}\tilde{A}}{\mathrm{d}\tau}+\frac{4\tilde{A}}{a}\frac{\mathrm{d}a}{\mathrm{d}\tau}=-\frac{\tilde{A}}{1-\tilde{A}},
        \label{b}
    \end{equation}
\end{subequations}
where $\tilde{A}(\tau=0)=1/b$ and $a(\tau=0)=1$. By substituting $\partial_\tau a/a = -1/4(1-\tilde{A})-\partial_\tau \tilde{A}/4\tilde{A}$ from (\ref{b}) into (\ref{a}), we find 
\begin{equation}
    \frac{6\beta}{1-\beta}\frac{1}{b}\left(\frac{1}{4(1-\tilde{A})}+\frac{1}{4\tilde{A}}\frac{\mathrm{d}\tilde{A}}{\mathrm{d}\tau}\right)+\frac{\tilde{A}}{1-\tilde{A}}=\frac{\tilde{Ec}}{a}.
\end{equation}
When $\tau = \mathcal{O}(1)$ and $b\gg 1$, we can obtain approximately  $a = \tilde{Ec}(\tilde{A}^{-1}-1)$ and substitute it into (\ref{b}), yielding 
\begin{equation}
    \frac{\mathrm{d}\tilde{A}}{\mathrm{d}\tau}=\frac{\tilde{A}}{3+\tilde{A}} \Rightarrow \tilde{A}(\tau)-\tilde{A}(\tau=\epsilon)+3\ln{\frac{\tilde{A}(\tau)}{\tilde{A}(\tau=\epsilon)}}=\tau-\epsilon
\end{equation}
where $\epsilon\sim b^{-1}\ll1$.
When considering $\tilde{\tau} = b\tau =\mathcal{O}(1)$, 
(\ref{a}) and (\ref{b}) are converted into
\begin{subequations}
    \begin{equation}
        \frac{6\beta}{1-\beta}\frac{1}{a}\frac{\mathrm{d}a}{\mathrm{d}\tilde{\tau}}+\frac{\tilde{Ec}}{a}=\frac{A_{zz}}{b-A_{zz}}
        \label{a_tidle}
    \end{equation}
    \begin{equation}
        \frac{\mathrm{d}A_{zz}}{\mathrm{d}\tilde{\tau}}+\frac{4A_{zz}}{a}\frac{\mathrm{d}a}{\mathrm{d}\tilde{\tau}}=-\underbrace{ \frac{A_{zz}}{b-A_{zz}}}_{A_{zz}\ll b, b\gg1} \approx 0
        \label{b_tidle}
    \end{equation}
\end{subequations}
By substituting $A_{zz}=a^{-4}$ from (\ref{b_tidle}) with the initial conditions of $A(0)=a(0)=1$ into (\ref{a_tidle}), we can obtain
\begin{equation}
    \frac{6\beta}{1-\beta}\frac{\mathrm{d}a}{\mathrm{d}\tilde{\tau}}=\frac{a}{a^4b-1}-\tilde{Ec}
\end{equation}
which can be transformed and integrated along with time, yielding 
\begin{equation}
    \tilde{\tau} (a)=\int_1^a\frac{6\beta}{1-\beta}\frac{(bx^4-1)}{x-x^4b\tilde{Ec}+\tilde{Ec}}dx
\end{equation}
This demonstrates that the full model (\ref{dimensionless_eq}) does not yield an explicit and straightforward approximate solution.

\subsubsection{2. Numerical results}
The governing equations (\ref{dimensionless_eq}) can be solved numerically. We solved the equations using MATLAB's \texttt{ode15s} solver, with absolute and relative tolerances set to $ 10^{-12}$. As an example, numerical results were obtained for $b = 10^4$, $\beta = 0.95$, and $Ec \in \{10^{2}, 10^{3}, 10^{4}\}$, {with the outcomes of dimensionless diameter $a(\tau)$, conformation $A_{zz}$ and $A_{rr}$, $\lambda_e/\lambda=\mathrm{log}_{10}(2/3E)$ and elastic-viscous stress ratio $\xi=\sigma_p/\sigma_s$, as presented in Figs.~S10(a-e)}. The onset of the EC phase was determined by calculating the maximum value of $2/3E$ when it is less than 1, as this corresponds to the point where the slope of the $a(\tau)$ curve is minimized. The solid markers in the figure represent the points at which $\lambda_e/\lambda$ were determined numerically.

Subsequently, we varied the conditions across $\beta \in \{0.05:0.05:0.95\}$, $b \in \{10^2:10:10^6\}$, and $Ec \in \{10^{1}:10:10^{5}\}$ to calculate the variation of $\lambda_e / \lambda$. The results (Fig.~S10(f)) indicate that $\lambda_e / \lambda$ is primarily influenced by $Ec$ and $b$, with negligible dependence on $\beta$ when $\beta\gtrsim 0.8$. This observation allows us to construct a two-dimensional phase diagram of $\lambda_e / \lambda$ as a function of $Ec$ and $b$, as shown in Fig.~S10(g), which intuitively illustrates that smaller $Ec$ and larger $b$ result in $\lambda_e$ being closer to the true relaxation time $\lambda$. However, considering that $Ec$ contains $\lambda$, which cannot be directly measured experimentally, we transform the phase diagram into a plot of $Ec_e$ and $b$ (Fig.~4 in the main text).

\subsubsection{3. Determine actually relaxation time}
Following the comparison of $\lambda_e/\lambda_{e,min}$ and $\lambda_e/\lambda$ between experimental and numerical results, as introduced in the main text, one can obtain the actual relaxation time $\lambda$ and extensibility $b$ as shown in Fig.~3 for PEO-PEG-8M and Fig.~\ref{fig:S11} for PEOvis. 

\begin{figure*}[b]
\includegraphics[width=0.85\textwidth]{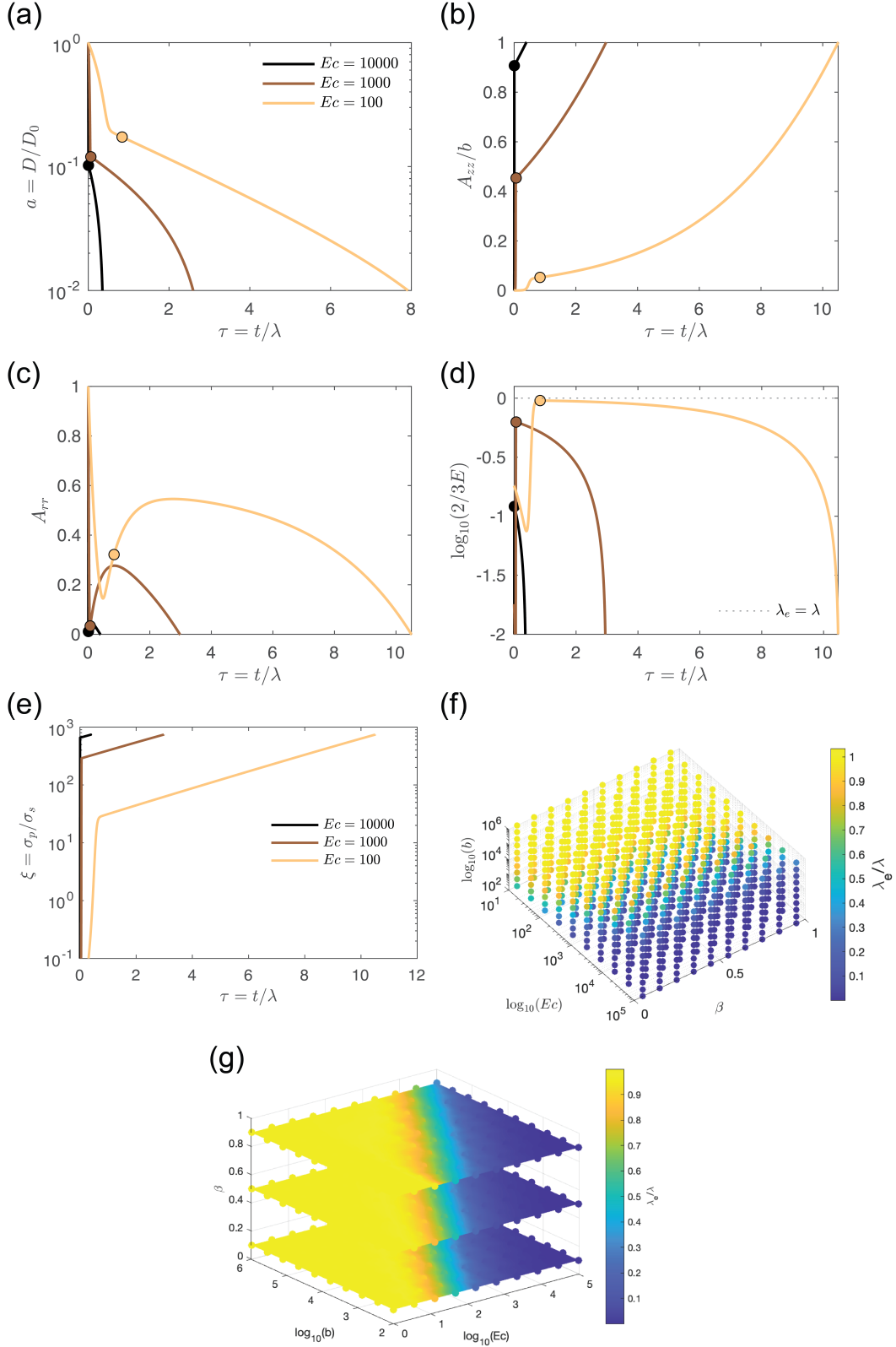}
\caption{\label{fig:FigureS10}  
Numerical results of the full model. Variations of (a) dimensionless diameter $a$, (b) normalized conformation component $A_{zz}/b$ and (c) conformation component $A_{rr}$, and $2/3E$ versus dimensionless time $\tau$, under the conditions of $\beta=0.95$, $b=10^4$ and $Ec=2\gamma\lambda/(D_0\eta_p) \in \{10^{2},10^{3},10^{4}\}$. The solid symbols represent the point we can obtain the $\lambda_e/\lambda=2/3E$ from the slope of $a(\tau)$ based on the numerical results. 
{(e) Evolution of elastic-viscous stress ratio $\xi=\sigma_p/\sigma_s$ during breakup for $Ec\in\{10^2,10^3,10^4\}$}
(f) Comparison of $\lambda_e/\lambda$ under the varying conditions of $\beta\in\{0.05:0.05:0.95\}$, $b\in\{10^2:10:10^6\}$ and $Ec\in\{10^{1}:10:10^{5}\}$. (g) Phase map of $\lambda_e/\lambda$ along with $b$ and $Ec$ at specific $\beta\in\{0.1, 0.5, 0.9\}$.
}
\label{fig:S10}
\end{figure*}

\begin{figure*}
\includegraphics[width=\textwidth]{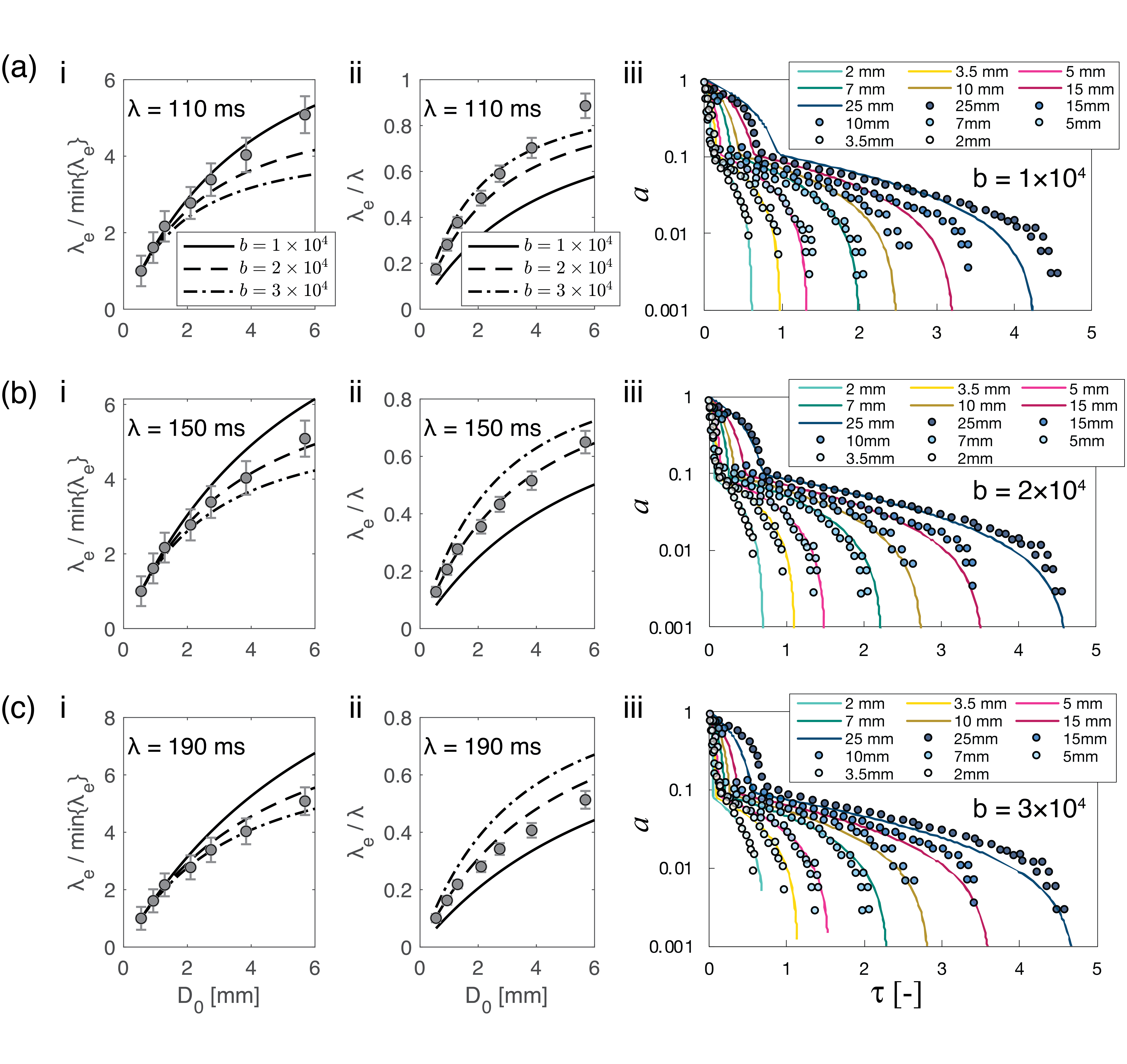}
\caption{\label{fig:FigureS11}  
Predicted results based on the procedure in Section IV-C3 using $\lambda = 110$ ms (a-i,ii), $\lambda = 150$ ms (b-i,ii), and $\lambda = 190$ ms (c-i,ii). 
Panels (a-iii, b-iii, c-iii) show the corresponding comparisons with $a(\tau)$ for the dataset from Ref.~\cite{Gaillard_2024}. 
The results suggest that $(\lambda, b) = (150~\mathrm{ms}, 2\times10^4)$ provides a better estimate compared to $(\lambda_{e,\max} = 110~\mathrm{ms}, b = 2\times10^4)$, as proposed in Ref.~\cite{Gaillard_2024}, which was based on the maximum measured $\lambda_{e,\max}$.
}
\label{fig:S11}
\end{figure*}

\section{V. Estimation of filament diameter $D_0$ in breakup for dripping and DoS method}
In the dripping and DoS methods, liquid breakup is triggered by the stretching of the liquid column due to the droplet’s own weight or spreading on a solid surface. Consequently, the initial minimum diameter of the suspended droplet, defined by the nozzle outer diameter \(\ell\), does not correspond to the initial diameter \(D_0\) governing the breakup dynamics. In contrast, in the stepwise-controlled stretching between two parallel plates, the initial breakup diameter \(D_0\) can be directly observed \cite{Gaillard_2024}. 
\begin{figure*}
    \centering
\includegraphics[width=\linewidth]{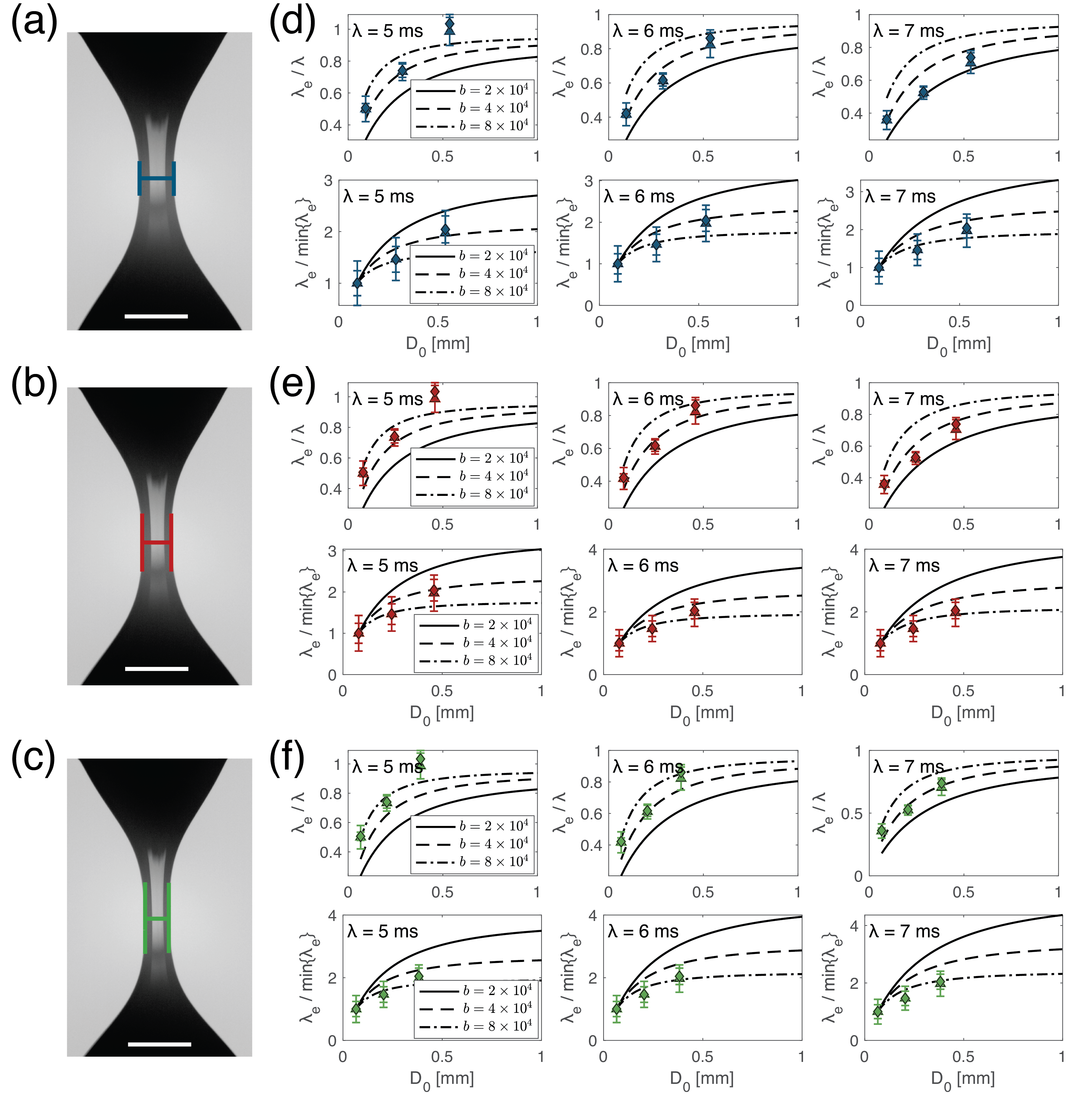}
    \caption{Comparison of different choices of $D_0$ in extracting the actual relaxation time $\lambda$. 
Based on conditions determined from aspect ratios $L/D \approx (a)~1$, (b)~2, and (c)~3, we evaluate $\lambda_e/\lambda_{e,\min}$ and $\lambda_e/\lambda$ following the same procedure. 
The corresponding results, shown in (d–f), indicate that selecting the aspect ratio within $1 \lesssim L/D \lesssim 3$ provides a reliable criterion. 
The white scale bar in (a–c) represents 1 mm.
}
    \label{FigureS12}
\end{figure*}
Therefore, in this study, we determine the initial breakup diameter $D_0$ based on the fundamental assumption of the full model, which considers a uniform liquid column. Through image recognition processing, we extract the liquid column diameter \(D\) and approximate length \(L\), and define the initial breakup diameter when the aspect ratio \(L/D\) satisfies \(1 \lesssim L/D \lesssim 3\). Under this assumption, we examine the influence of selecting \(L/D \approx 1\), \(2\), and \(3\) on the extraction of the actual relaxation time, as shown in Fig.~\ref{FigureS12} by taking PEO-PEG-8M for example. It is observed that within this range of \(D_0\), the conclusions regarding \(\lambda\) and \(b\) remain unaffected, corroborating the robustness of our proposed criterion.


%